\pdfoutput=1

\documentclass{aastex}

\usepackage{graphicx,amssymb}
\usepackage{epsf,psfig}
\usepackage{txfonts}
\usepackage{multirow}
\usepackage{bigstrut}

\received{2012 January 6}
\accepted{2012 February 28}

\begin{document}

\title{ORDER AND CHAOS IN A THREE-DIMENSIONAL BINARY SYSTEM OF INTERACTING GALAXIES}

\author{Euaggelos E. Zotos}

\affil{Department of Physics, \\
Section of Astrophysics, Astronomy and Mechanics, \\
Aristotle University of Thessaloniki \\
GR-541 24, Thessaloniki, Greece}
\email{evzotos@astro.auth.gr}

\shorttitle{Order and chaos in a 3D binary system of interacting galaxies}
\shortauthors{Euaggelos E. Zotos}

\begin{abstract}

In the present article, we present a galactic gravitational model of three degrees of freedom, in order to investigate and reveal the behavior of orbits in a binary quasar system. The two quasars are hosted in a pair of interacting disk galaxies. We study in detail the regular or chaotic character of motion, in two different cases: the time independent model in both 2D and 3D dynamical systems and the evolving 3D model. Our numerical calculations, indicate that a large fraction of orbits in the 2D system are chaotic in the time independent case. A careful analysis suggest that several Lindblad resonances are also responsible for the chaotic motion of stars in both hosts galaxies. In the time dependent system, we follow the evolution of 3D orbits in our dynamical model, as the two interacting host galaxies develop dense and massive quasars in their cores, by mass transportation from the disks to their nuclei. In this interesting case, there are orbits that change their orbital character from regular to chaotic and vise versa and also orbits that maintain their character during the galactic evolution. These results strongly indicate that the ordered or chaotic nature of 3D orbits depends, not only in the galactic interaction but also in the presence of the quasars in the galactic cores of the host galaxies. The outcomes derived from our dynamical model are compared with observational data. Some theoretical arguments to support the numerically derived outcomes are presented, both in 2D and 3D systems, while a comparison with similar earlier work is also made.

\end{abstract}


\section{INTRODUCTION}

Quasars were first discovered by Maarten Schmidt in 1963 as very distant, highly energetic stellar objects with a star like appearance. Since the discovery of the first quasar, astronomers have been trying to reveal and understand the conditions of birth and existence of these immensely powerful objects. Today, it is believed that quasars are the extremely luminous cores at the centers of distant galaxies, surrounding supermassive black holes (SMBHs). Binary quasars are pairs of quasars bound together by gravitational forces. Binary quasars, like other quasars, are thought to be the products of galaxy mergers. Recent observations indicate that a large number of bright quasars are hosted in massive spiral galaxies with really prominent disks (Letawe et al., 2006). In an early paper (Papadopoulos and Caranicolas, 2005), a simple dynamical model with a rotating disk and a massive nucleus was presented, in order to study the properties of motion in a quasar. Furthermore, there is strong evidence (Letawe et al., 2007) that interacting galaxies host powerful quasars with strong emission lines. It is also found that about $50\%$ of the observed host galaxies display signs of interaction, which are consistent with previous studies (see Hutchings \& Neff, 1992; Bahcall et al., 1997; Boyce et al., 1999; Kaspi et al., 2000; Floyd et al., 2004; Jahnke et al., 2004). Because most such mergers would have occurred in the very distant past, binary quasars and their associated galaxies, are very away and therefore, difficult for most telescopes to resolve.

Today, the only well known close binary pair of SMBHs is OJ 287 (see Sillanp\"{a}\"{a} et al., 1988 and also later papers on OJ 287). Sillanp\"{a}\"{a} et al. (1988) conducted numerical experiments, using an $N$-body simulation code in order to study the mass flows. These simulations revealed that the inflow into the center of a black hole will produce an outburst. Tidal interaction and thus the luminosity of the binary system, would be greatest in the last stages which is consistent with OJ 287 being one of the brightest quasars. Moreover, under the tidal triggering hypothesis, binary galaxies with morphological signs of interaction in their disks, should be more likely to display Seyfert or quasar activity in their cores (see Byrd et al. 1986; Sundelius et al. 1987 and references therein about accretion in interacting galaxies).

Therefore, it seems very challenging to construct a gravitational model of three degrees of freedom (3D), in order to study the dynamical properties in a pair of interacting disk galaxies hosting quasars. The present article is organized as follows: In Section 2 we present our gravitational dynamical model, which describes the motion in binary quasar stellar system. In Section 3 we provide an analysis of the 2D system, considering orbits in the galactic plane $(z=0)$. In the next Section, we study the character of motion in the 3D system, using different kinds of dynamical methods. Some interesting semi-numerical results are also provided in the same Section. In Section 5 we use a 3D time dependent model, in order to follow the evolution of orbits as the host galaxies develop massive and dense quasars in their centers. We conclude with Section 6, where the discussion is presented and a comparison between the present theoretical results with observational data is made.

\section{PRESENTATION OF THE DYNAMICAL MODEL}

Our gravitational model consists of a pair of disk galaxies, each having a dense, massive and spherically symmetric nucleus. The potential which describes the motion in the first host galaxy (hereafter galaxy G1) is given by the equation
\begin{equation}
V_1(r,z) = V_{n1}(r,z) + V_{d1}(r,z) = - \frac{M_{n1}}{\sqrt{r^2+z^2+c_{n1}^2}}
- \frac{M_{d1}}{\sqrt{b_1^2+r^2+\left(a_1+\sqrt{h_1^2+z^2}\right)^2}} \ \ \ ,
\end{equation}
where $r^2=x^2+y^2$, $M_{n1},M_{d1}$ is the mass of the nucleus and the disk of galaxy 1 respectively, $a_1$ is the disk's scale length, $h_1$ is the disk's scale height, $b_1$ is the core radius of the disk halo, while $c_{n1}$ is the scale length of the nucleus. The second host galaxy (hereafter galaxy G2) is described by the potential
\begin{equation}
V_2(r,z) = V_{n2}(r,z) + V_{d2}(r,z) = - \frac{M_{n2}}{\sqrt{r^2+z^2+c_{n2}^2}}
- \frac{M_{d2}}{\sqrt{b_2^2+r^2+\left(a_2+\sqrt{h_2^2+z^2}\right)^2}} \ \ \ ,
\end{equation}
where again $r^2=x^2+y^2$, $M_{n2},M_{d2}$ is the mass of the nucleus and the disk of galaxy 2 respectively, $a_2$ is the disk's scale length, $h_2$ is the disk's scale height, $b_2$ is the core radius of the disk halo, while $c_{n2}$ is the scale length of the nucleus. The disks of the two host galaxies are represented by the well known Miyamoto-Nagai model (Miyamoto \& Nagai, 1975). The Plummer sphere we choose to describe each nucleus, has been used many times in the past, in order to study the effects of the introduction of a central mass component in the core of a galaxy (see Hasan and Norman, 1990; Hasan et al., 1993).

In our study we shall use the theory of the circular restricted three body problem (see Caranicolas and Inanen, 2009; Caranicolas and Papadopoulos, 2009; Caranicolas and Zotos, 2009). The two bodies move in circular orbits in an inertial frame OXYZ, with the origin at the center of mass of the system, with a constant angular frequency $\Omega_p > 0$, given by Kepler's third law
\begin{equation}
\Omega_p= \sqrt{\frac{G M_t}{R^3}} \ \ \ ,
\end{equation}
where $M_t=M_{n1}+M_{d1}+M_{n2}+M_{d2}$ is the total mass of the system, while $R$ is the distance between the centers of the two galaxies. A clockwise, rotating frame Oxyz, is used with axis Oz coinciding with the axis OZ and the axis Ox coinciding with the straight line joining the two bodies. In this frame, which rotates with angular frequency $\Omega_p$, the two galactic centers have fixed positions $C_1(x,y,z) = \left(x_1,0,0\right)$ and $C_2(x,y,z) = \left(x_2,0,0\right)$ respectively. The total potential which is responsible for the motion of a star in the dynamical system of the binary quasar is
\begin{equation}
\Phi_t(x,y,z)=\Phi_{G1}(x,y,z) + \Phi_{G2}(x,y,z) + \Phi_{rot}(x,y,z) \ \ \ ,
\end{equation}
where
\begin{eqnarray}
\Phi_{G1}(x,y,z) &=& - \frac{M_{n1}}{\sqrt{r_1^2+c_{n1}^2}}
- \frac{M_{d1}}{\sqrt{b_1^2+r_{a1}^2+\left(a_1+\sqrt{h_1^2+z^2}\right)^2}} \ \ \ , \nonumber \\
\Phi_{G2}(x,y,z) &=& - \frac{M_{n2}}{\sqrt{r_2^2+c_{n2}^2}}
- \frac{M_{d2}}{\sqrt{b_2^2+r_{a2}^2+\left(a_2+\sqrt{h_2^2+z^2}\right)^2}} \ \ \ , \nonumber \\
\Phi_{rot}(x,y) &=& -\frac{\Omega_p^2}{2}\left[\frac{M_2}{M_t}r_{a2}^2 + R_s r_{a1}^2 -
R^2 \frac{M_2}{M_t}R_s \right] \ \ \ ,
\end{eqnarray}
and
\begin{eqnarray}
r_{a1}^2 &=& \left(x-x_1\right)^2 + y^2, \ \ \ r_{a2}^2 = \left(x-x_2\right)^2 + y^2 \ \ \ , \nonumber \\
r_1^2 &=& r_{a1}^2 + z^2, \ \ \ r_2^2  =  r_{a2}^2 + z^2 \ \ \ ,
\end{eqnarray}
with
\begin{eqnarray}
x_1 &=& -\frac{M_2}{M_t}R, \ \ \ x_2  =  R - \frac{M_2}{M_t}R = R + x_1 \ \ \ , \nonumber \\
M_2 &=& M_{n2}+M_{d2}, \ \ \ R_s = 1-\frac{M_2}{M_t} \ \ \ .
\end{eqnarray}

The angular frequency $\Omega_p$ is calculated as follows: The two bodies move around their common mass center of the system with angular frequencies $\Omega_{p1}$ and $\Omega_{p2}$, given by
\begin{eqnarray}
\Omega_{p1} &=& \sqrt{\frac{1}{x_1}\left(\frac{-d V_2(r)}{dr}\right)_{r=R}} \ \ \ , \nonumber \\
\Omega_{p2} &=& \sqrt{\frac{1}{x_2}\left(\frac{d V_1(r)}{dr}\right)_{r=R}} \ \ \ .
\end{eqnarray}
As the two bodies are not mass points the two angular frequencies are not equal, in general. However, this issue can easily be resolved. The angular frequencies of the two bodies, can become equal under reasonable assumptions. This is justified, if the final set of the parameters has physical meaning and represents satisfactorily the dynamical system. The equation $\Omega_{p1}=\Omega_{p2}$ leads to an eightfold infinity of solutions in the eight unknowns $\left(a_1, a_2, b_1, b_2, h_1, h_2, c_{n1}, c_{n2} \right)$. If one chooses proper values (representing the dynamical system), lets say for the seven parameters $\left(a_2, b_1, b_2, h_1, h_2, c_{n1}, c_{n2} \right)$, then equation $\Omega_{p1}=\Omega_{p2}$ gives only two values for the parameter $a_1$. One value is positive, while the other is negative and is rejected. The author would like to make clear that, after choosing properly the parameters, the deviation between the two angular frequencies is negligible, so that $\nu =
\left(| \Omega_{p1}-\Omega_{p2} |\right)/\Omega_{p1}$ is of order of $10^{-8}$ or even smaller and $\xi = | \Omega_{p1}-\Omega_p |$ or $ \xi = | \Omega_{p2}-\Omega_p |$ is of order of $10^{-6}$. Therefore, we consider the two angular frequencies practically equal, that is $\Omega_{p1}=\Omega_{p2}=\Omega_p$. Moreover, the treatment of large bodies with spherical symmetry as mass points is very common in celestial modeling. This method is quite familiar to those measuring the masses of disk galaxies. The fact that the two bodies (host galaxies), are sufficiently apart from each other, allow us to assume that the tidal phenomena are very small and therefore negligible.

In this rotating frame the equations of motion are
\begin{eqnarray}
\ddot{x} &=& -\frac{\partial \Phi_t}{\partial x} - 2 \Omega_p \dot{y} \ \ \ , \nonumber\\
\ddot{y} &=& -\frac{\partial \Phi_t}{\partial y} + 2 \Omega_p \dot{x} \ \ \ , \nonumber\\
\ddot{z} &=& -\frac{\partial \Phi_t}{\partial z} \ \ \ ,
\end{eqnarray}
where the dot indicates derivative with respect to the time. The only integral of motion for the system of differential equations (9),
is the Jacobi integral given by the equation
\begin{equation}
J=\frac{1}{2}\left(p_x^2 + p_y^2 + p_z^2 \right) + \Phi_t(x,y,z) = E_J \ \ \ ,
\end{equation}
where $p_x, p_y, p_z$ are the momenta per unit mass conjugate to $x,y,z$, while $E_J$ is the numerical value of the Jacobi integral.

All numerical outcomes of the present work, are based on the numerical integration of the equations of motion (9), which was made using a Bulirsh-St\"{o}er routine in Fortran 95, with double precision in all subroutines. The accuracy of the calculations was checked by the consistency of the Jacobi integral (10), which was conserved up to the eighteenth significant figure.

In this article, we shall use a system of galactic units, where the unit of length is 20 kpc, the unit of mass is $1.8 \times 10^{11} M_\odot$ and the unit of time is $0.99 \times 10^8$ yr. The velocity unit is 197 km/s, while $G$ is equal to unity (see Vozikis \& Caranicolas, 1992). In these units, we use the values: $a_1=0.15, b_1=0.2542, h_1=0.00925, c_{n1}=0.0125, a_2=0.175, b_2=0.0789, h_2=0.00875, c_{n2}=0.01$. The values of the above quantities of the dynamical system remain constant during this research, while the values of $M_{n1}, M_{d1}, M_{n2}, M_{d2}$ and $R$ are treated as parameters. The above numerical values of the constant dynamical quantities of the system, secure positive density everywhere and free of singularities.
\begin{figure}[!tH]
\centering
\resizebox{0.9\hsize}{!}{\rotatebox{0}{\includegraphics*{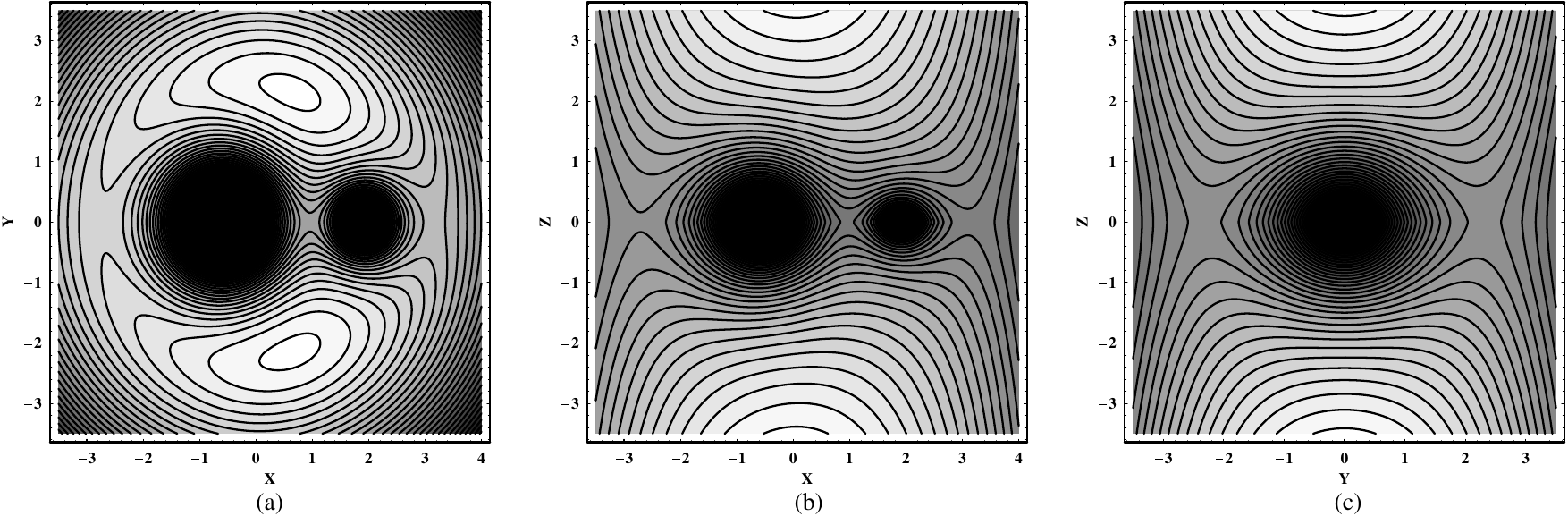}}}
\caption{(a-c): Contours of the projections of the iso-potential curves $\Phi_t(x,y,z) = E_J$ on the $(x,y)$, $(x,z)$ and $(y,z)$ planes.}
\end{figure}
\begin{figure}[!tH]
\centering
\resizebox{0.50\hsize}{!}{\rotatebox{0}{\includegraphics*{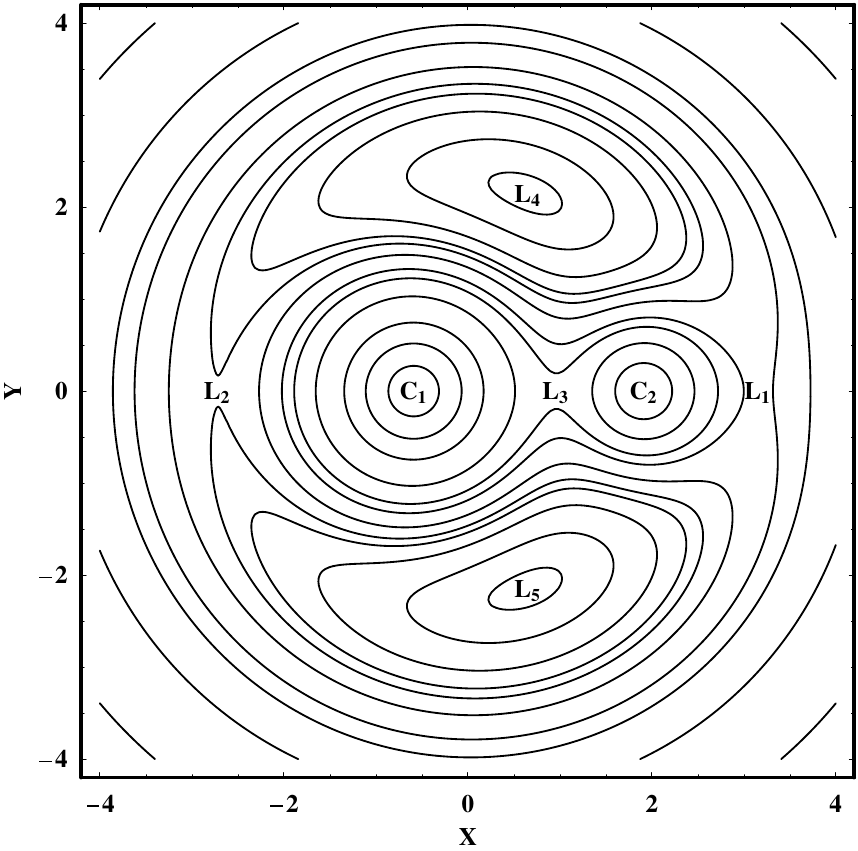}}}
\caption{Contours of the projections of the iso-potential curves $\Phi_t(x,y,z) = E_J$ on the $(x,y)$ plane. The five Lagrange equilibrium points are indicated as: $L_1, L_2, L_3, L_4$ and $L_5$, while $C_1$ and $C_2$ are the centers of the two host galaxies at a distance $R=2.5$.}
\end{figure}

Figure 1a-c shows the contours of the projections of the iso-potential curves $\Phi_t(x,y,z)=E_J$, on the $(x,y), (x,z), (y,z)$ planes respectively. The values of the parameters are: $M_{n1}=0.08, M_{d1}=2.0, M_{n2}=0.04, M_{d2}=0.6, R=2.5$ and $\Omega_p=0.417229$. Lighter colors indicate higher values of $E_J$. Figure 2 shows the contours of the iso-potential curves $\Phi_t(x,y,z)=E_J$, on the $(x,y)$ plane, with some additional details regarding the stability and the structure of the dynamical system. Moreover, $L_1, L_2, L_3, L_4, L_5$ are the five Lagrange equilibrium points, while $C_1$ and $C_2$ are the centers of the two galaxies at a distance $R=2.5$. At these equilibrium points we have
\begin{equation}
\frac{\partial \Phi_t}{\partial x} = \frac{\partial \Phi_t}{\partial y} = \frac{\partial \Phi_t}{\partial z} = 0 \ \ \ .
\end{equation}
$L_1, L_2, L_3$ are the unstable saddle equilibrium points, while $L_4$ and $L_5$ are the triangular points (see Binney \& Tremaine, 2008).

\section{STRUCTURE OF THE 2D HAMILTONIAN SYSTEM}

In this section we shall investigate the properties of motion in the Hamiltonian system of two degrees of freedom (2D). This can be derived from equation (10), if we set $z=p_z=0$. Then the corresponding Hamiltonian writes
\begin{equation}
J_2=\frac{1}{2}\left(p_x^2 + p_y^2 \right) + \Phi_t(x,y) = E_{J2} \ \ \ ,
\end{equation}
where $E_{J2}$ is the numerical value of $J_2$. As the dynamical system is now 2-dimensional, we can use the classical method of the $x-p_x, y=0, p_y>0$ Poincar\'{e} surface of section, in order to determine the regular or chaotic character of motion. The results obtained from the study of the 2D system, will be used in order to help us to understand the structure of the more complicated phase space of the 3D system, which will be presented in the following section.
\begin{figure}[!tH]
\centering
\resizebox{0.9\hsize}{!}{\rotatebox{0}{\includegraphics*{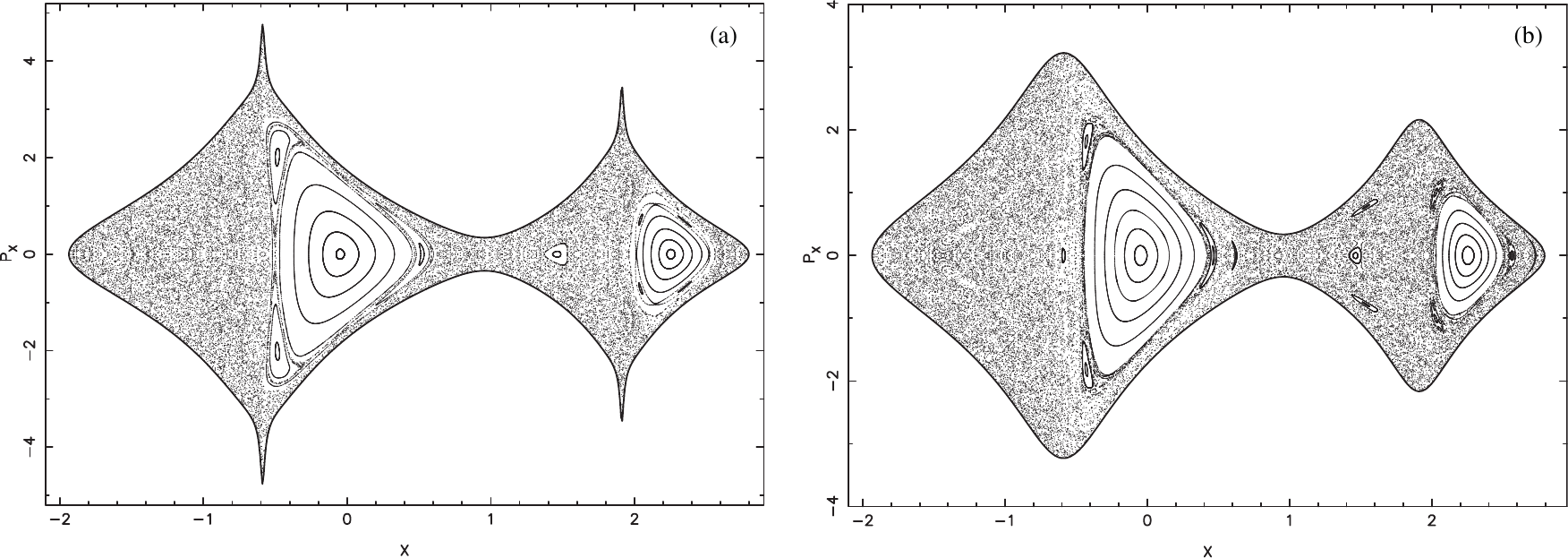}}}
\caption{(a-b): The $(x,p_x)$ Poincar\'{e} phase planes when $R=2.5, E_{J2}=-2.0$ and $\Omega_p=0.417229$.}
\end{figure}
\begin{figure}[!tH]
\centering
\resizebox{0.9\hsize}{!}{\rotatebox{0}{\includegraphics*{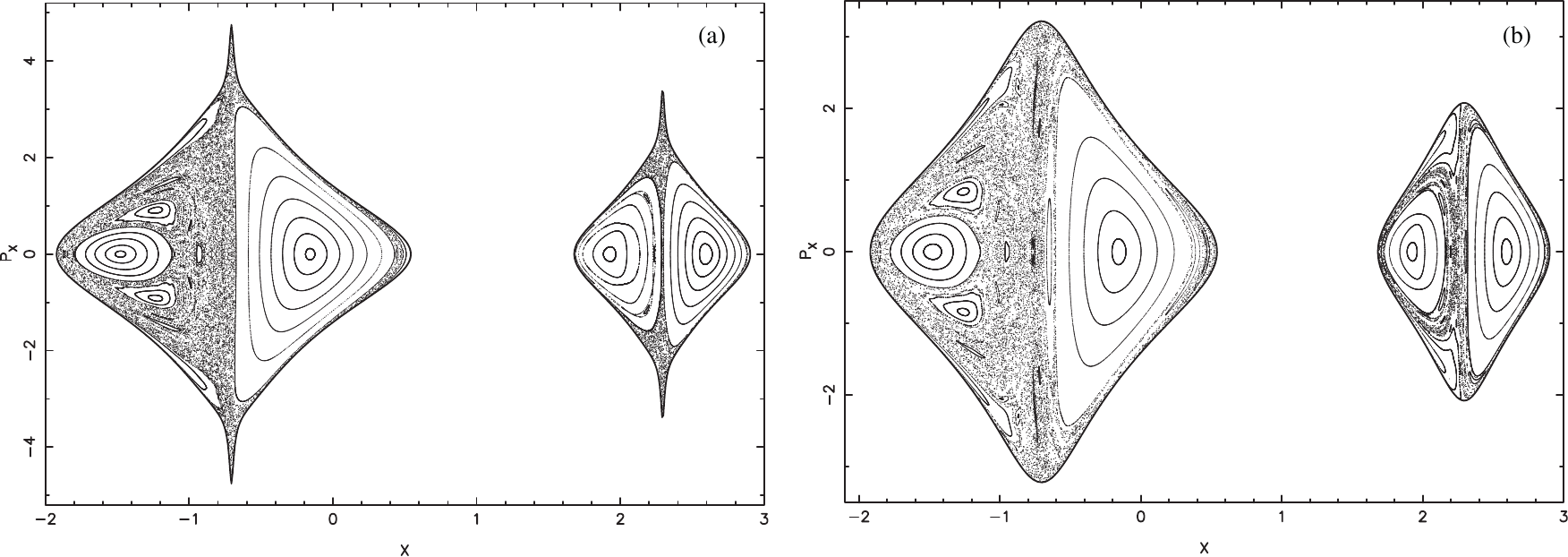}}}
\caption{(a-b): Similar as in Fig. 3a-b when $R=3.0, E_{J2}=-2.0$ and $\Omega_p=0.317397$.}
\end{figure}
\begin{figure}[!tH]
\centering
\resizebox{0.9\hsize}{!}{\rotatebox{0}{\includegraphics*{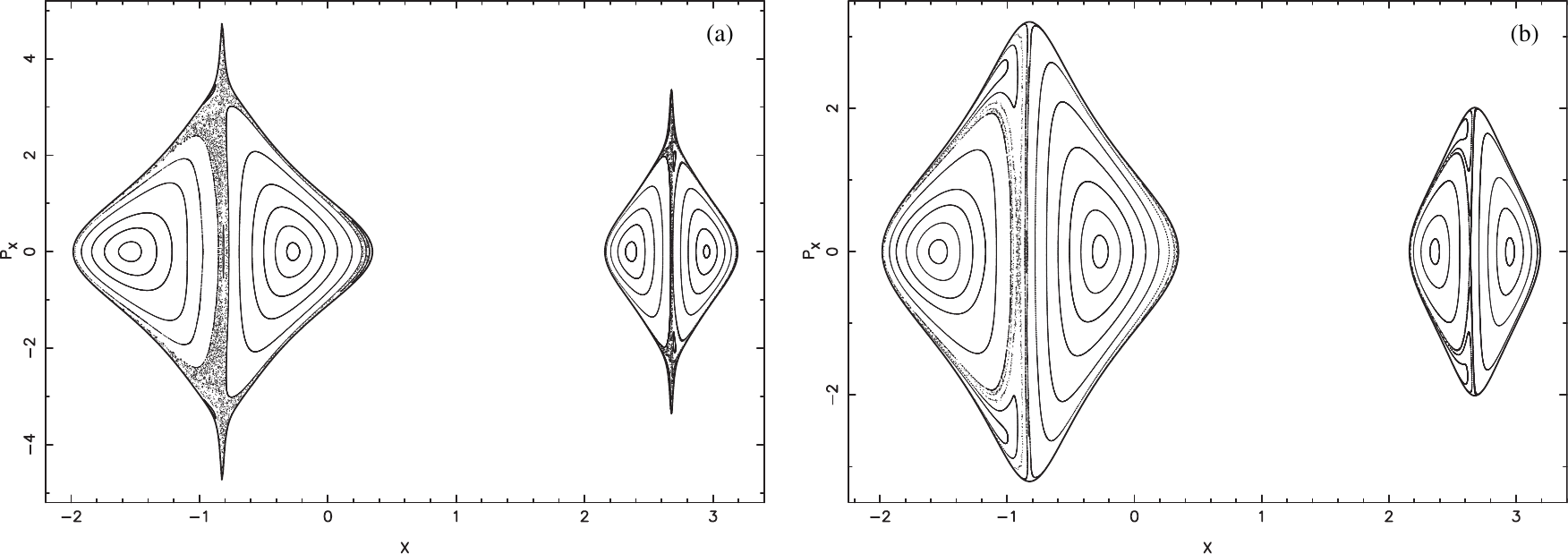}}}
\caption{(a-b): Similar as in Fig. 3a-b when $R=3.5, E_{J2}=-2.0$ and $\Omega_p=0.251830$.}
\end{figure}

Figure 3a shows the $(x,p_x)$ phase plane, when the distance between the centers of the two galaxies is $R=2.5$. The values of the other parameters are: $M_{n1}=0.08, M_{d1}=2.0, M_{n2}=0.04, M_{d2}=0.6, \Omega_p=0.417229$. The value of the Jacobi integral is $E_{J2}=-2.0$. We observe that the regular motion is confined around the stable retrograde periodic point, in the galaxy 1 (G1), while in the galaxy 2 (G2) there are regular regions around both, the direct (i.e. in the same direction as the rotation) and the retrograde periodic points. The rest of the phase plane is occupied by a large unified chaotic sea surrounding both host galaxies. There are also some small islands, corresponding to secondary resonances. Another interesting point in Fig. 3a, is the high velocities, observed near the the centers of the two host galaxies. This results from the presence of the dense nucleus in each galactic core and it is characteristic of nuclear galactic activity. Figure 3b is similar to Fig. 3a, but when $M_{d1}=2.08$ and $M_{d2}=0.64$. As the total mass of the system is conserved, this means that now the mass is on the disks of the two galaxies. In this case, the pattern has one main difference from the pattern shown in Fig. 3a. Here the chaotic regions are low near the center of each galaxy and the velocities near the two galactic cores are smaller compared to those of Fig. 3a. Therefore, one can say, that the structure of chaos in both galaxies is not only a result of galactic interaction, but it is also a result of nuclear galactic activity. In other words, the presence of the quasars in each galactic core affects drastically the character of motion in the host galaxies. Moreover, the absence of the quasars reduces the velocities near the centers of the two host galaxies.

Figure 4a is similar to Fig. 3a but when the distance between the centers of two the galaxies is $R=3$ and $\Omega_p=0.317397$. In this case the two galaxies do not communicate directly in the $(x,p_x)$ phase plane and therefore, we can observe two separate regions of motion around each galaxy. The majority of orbits around the direct and retrograde periodic points in each galaxy is ordered. There is a considerable chaotic region, mainly near the central regions of each host galaxy. High velocities are once more observed, near the galactic cores of the two host galaxies, while some secondary resonances are also present. Figure 4b is similar to Fig. 3b but when the two host galaxies (G1 and G2) are quiet. Here, the secondary resonances look more prominent. Furthermore, as the quasars are absent, the velocities near the center of the two galactic cores are smaller than those observed in Fig. 4a.

Figure 5a is similar to Fig. 2a but when the distance between the centers of the two galaxies is $R=3.5$ and $\Omega_p=0.251830$. Once more, there are two separate regions of motion around each host galaxy. There is a relatively small chaotic layer, confined mainly near the central region of each galaxy, while the rest of the phase plane is covered by regular orbits circulating around the stable direct and retrograde periodic points. Some small sticky regions are also present in both galaxies. Figure 5b is similar to Fig. 3b, but when the two host galaxies (G1 and G2) are quiet. All orbits in both galaxies seem to be regular. Chaotic motion was not observed and, if present, is negligible. Therefore, we can say that, the two interacting host galaxies, for large separations, do not show chaotic motion when the quasars are not present in their galactic cores.

Our numerical results suggest that there are two kinds of chaotic orbits: (i) chaotic orbits approaching both galaxies and (ii) chaotic orbits that approach only one of the two host galaxies. Later in this section, we will see that whether a chaotic orbit approaches the two galaxies or only one of them, strongly depends on the integration time of the particular orbit. On the other hand, the ordered orbits circulate around only one of the two host galaxies. It would be of particular interest, to find regular orbits that could circulate around both host galaxies. Unfortunately, this kind of ordered orbits does not exist in our gravitational model, which describes a system of two interacting galaxies that host quasars in their cores.
\begin{figure}[!tH]
\centering
\resizebox{0.85\hsize}{!}{\rotatebox{0}{\includegraphics*{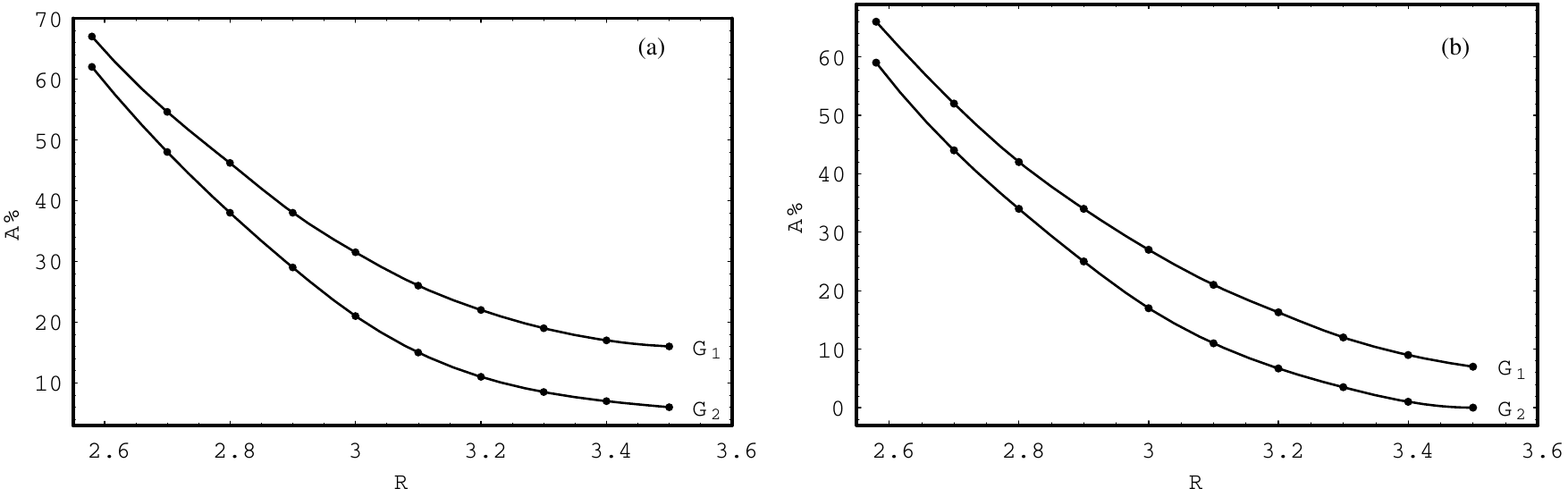}}}
\caption{(a-b): A plot of the area $A\%$ of the $(x,p_x)$ phase planes covered by chaotic orbits vs. distance $R$ for (a-left): the two active host galaxies and (b-right): the two quiet host galaxies.}
\end{figure}
\begin{figure}[!tH]
\centering
\resizebox{0.85\hsize}{!}{\rotatebox{0}{\includegraphics*{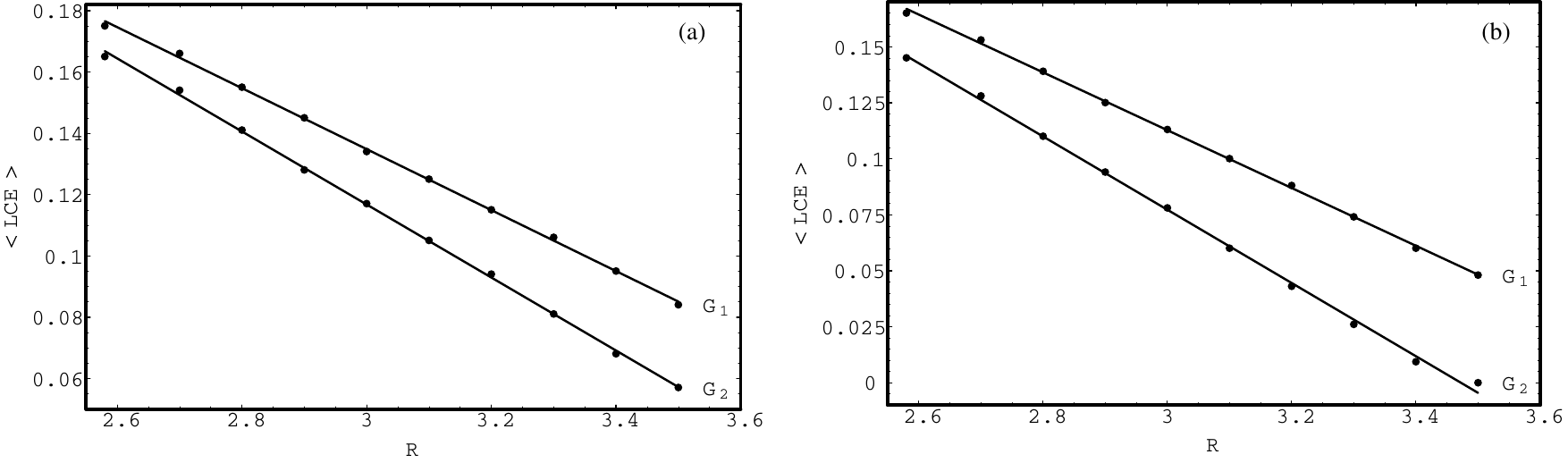}}}
\caption{(a-b): A plot of the average value of the maximum Lyapunov Characteristic Exponent $< LCE >$ vs. distance $R$ for (a-left): the two active host galaxies and (b-right): the two quiet host galaxies.}
\end{figure}
\begin{figure}[!tH]
\centering
\resizebox{0.8\hsize}{!}{\rotatebox{0}{\includegraphics*{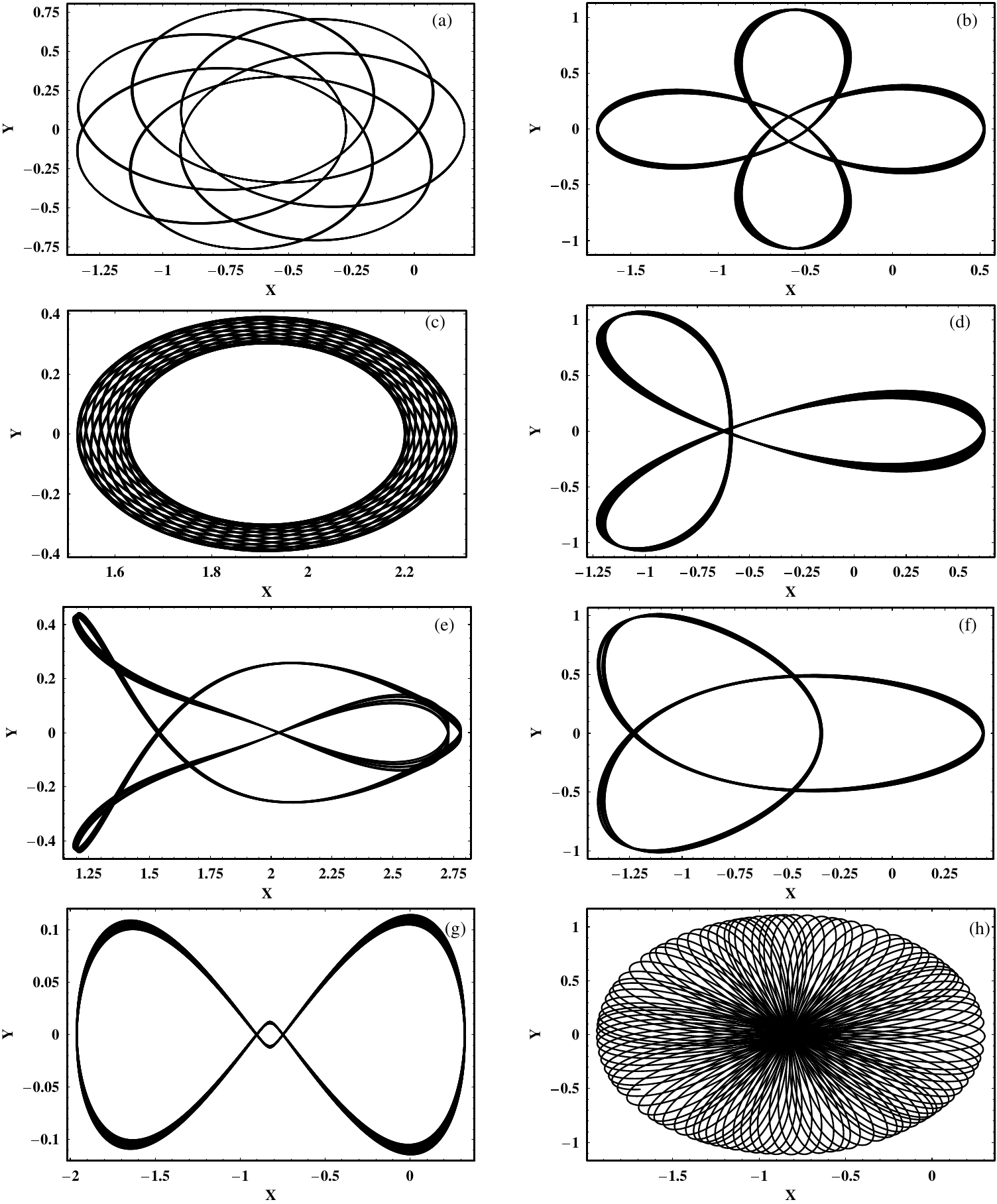}}}
\caption{(a-h): Eight representative regular orbits of the 2D dynamical system. The values of the initial conditions and all the other parameters are given in the text.}
\end{figure}
\begin{figure}[!tH]
\centering
\resizebox{0.8\hsize}{!}{\rotatebox{0}{\includegraphics*{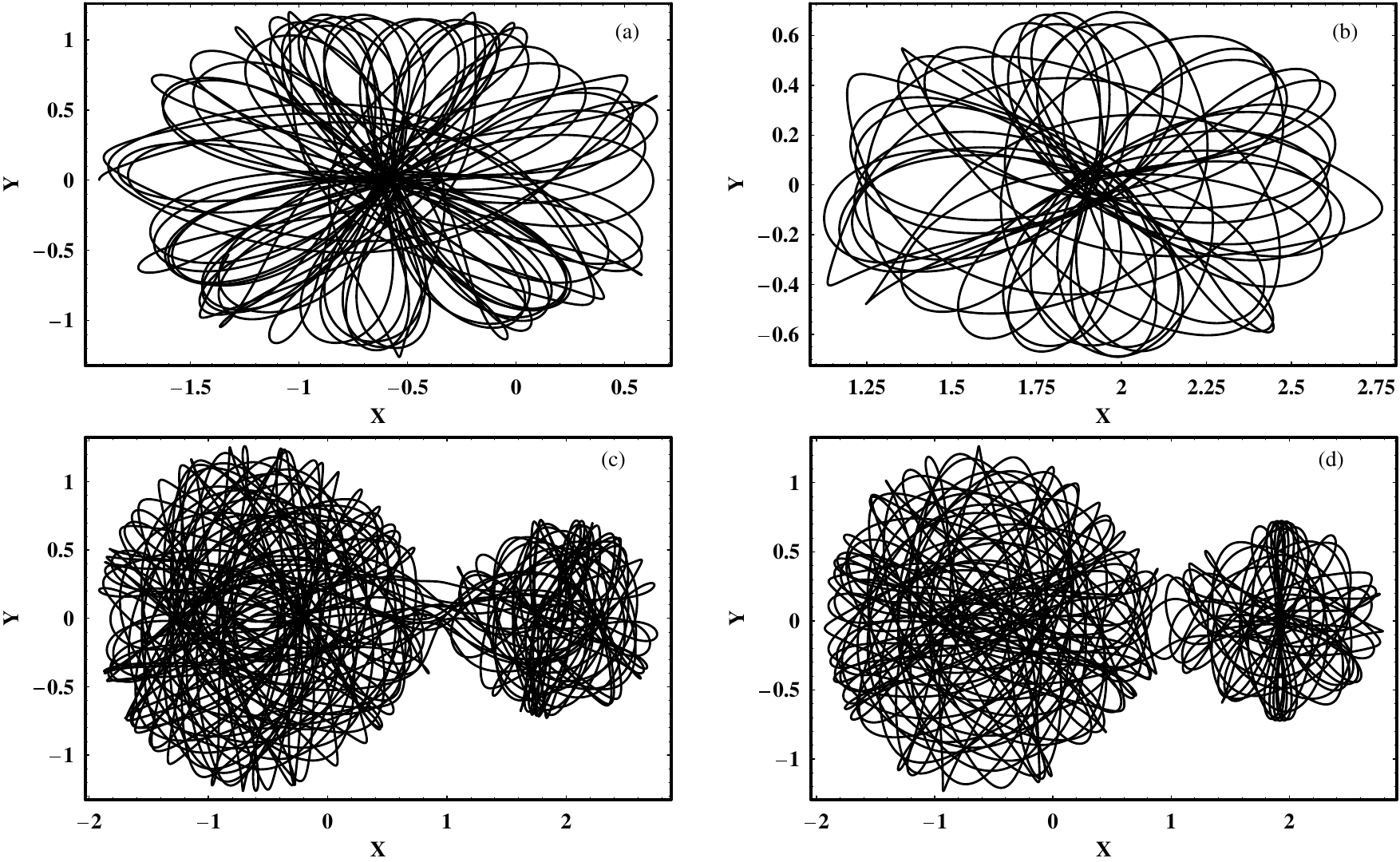}}}
\caption{(a-d): Four different chaotic orbits of the 2D dynamical system. See text for details.}
\end{figure}

Figure 6a shows the percentage $A\%$ of the $(x,p_x)$ phase planes covered by chaotic orbits as a function of the distance between the centers of the two active host galaxies. The values of the parameters are: $M_{n1}=0.08, M_{d1}=2.0, M_{n2}=0.04, M_{d2}=0.6$, while the value of the Jacobi integral is $E_{J2}=-2.0$. In this case, the quasars are present in the cores of two host galaxies G1 and G2. The value of the angular frequency $\Omega_p$ is calculated for each particular value of the distance $R$ from equation (3). The range of values regarding the distance between the centers of the two galaxies is $2.58 \leq R \leq 3.5$. This particular range was chosen so as the two galaxies do not communicate with each other directly, producing two separate $(x,p_x)$ phase planes, in order to be able to calculate the chaotic percentage of each phase plane more easily. The lower limit of the above range, is the minimum distance between the centers of the two galaxies, for which we have two separate $(x,p_x)$ phase planes as those shown in Figs. 4a and 5a. We observe from Fig. 6a that the percentage $A\%$ decreases exponentially, as the distance $R$ increases, for both active galaxies G1 and G2. Figure 6b is similar to Fig. 6a but for the case when the quasars are not present in the cores of the two galaxies. The values of the parameters are: $M_{n1}=0, M_{d1}=2.08, M_{n2}=0, M_{d2}=0.64$, while the value of the Jacobi integral is $E_{J2}=-2.0$. Here again the the percentage $A\%$ decreases exponentially, as the distance $R$ increases, for both quiet galaxies G1 and G2. A more detailed view of Figs. 6a and 6b reveals that for each particular value of the distance $R$, the chaotic percentage $A\%$ is always smaller when the quasars are not present, in both host galaxies G1 and G2. We must point out, that the percentage $A\%$ is calculated as follows: we choose 1000 orbits with different and random initial conditions $(x_0,p_{x0})$ in each phase plane and then divide the number of those who produce chaotic orbits to the total number of orbits.

Figure 7a shows a plot of the average value of the maximum Lyapunov Characteristic Exponent - LCE (see Lichtenberg and Lieberman, 1992) as a function of the distance between the centers of the two active host galaxies. The values of the parameters are: $M_{n1}=0.08, M_{d1}=2.0, M_{n2}=0.04, M_{d2}=0.6$, while the value of the Jacobi integral is $E_{J2}=-2.0$. In this case, the quasars are present in the cores of two host galaxies G1 and G2. The value of the angular frequency $\Omega_p$ is calculated for each particular value of the distance $R$ from equation (3). One can see, in Fig. 7a, that $< LCE >$ decreases linearly as the distance $R$ increases for both cases (G1 and G2). Figure 7b is similar to Fig. 7a but for the case when the quasars are not present in the cores of the two galaxies. The values of the parameters are: $M_{n1}=0, M_{d1}=2.08, M_{n2}=0, M_{d2}=0.64$, while the value of the Jacobi integral is $E_{J2}=-2.0$. Here again $< LCE >$ decreases linearly, as the distance $R$ increases, for both quiet galaxies G1 and G2. Once more, as we have pointed out in Figs. 6a and 6b, for each particular value of the distance $R$, the average value $< LCE >$ is always smaller when the quasars are not present, for both galaxies G1 and G2. Here we must note, that it is well known that the value of LCE is different in each chaotic component (see Saito \& Ichimura, 1979). As we have in all cases regular regions and only one unified chaotic sea in each $(x,p_x)$ phase plane, we calculate the average value of LCE by taking 500 orbits with different and random initial conditions $(x_0,p_{x0})$ in the chaotic sea in each case. Note that, all calculated LCEs were different on the fifth decimal point in the same chaotic region.

Figure 8a-h shows eight representative regular orbits of 2D dynamical system. Figure 8a shows an orbit circulating around host galaxy 1, with initial conditions: $x_0=0.2, y_0=0, p_{x0}=0$, while the value of $p_{y0}$ is obtained from the Jacobi integral (12) for all orbits. The value of Jacobi integral is $E_{J2}=-2.0$ always. The values of all the other parameters are as in Fig. 3a. Figure 8b shows a quasi periodic orbit moving around galaxy 1, with initial conditions: $x_0=0.52, y_0=0, p_{x0}=0$, while the values of all other parameters are as in Fig. 3a. This orbit is characteristic of the 3:3 resonance. In Figure 8c an ordered orbit circulating around galaxy 2 is shown. The initial conditions are: $x_0=2.2, y_0=0, p_{x0}=0$, while the value of all other parameters are as in Fig. 3a. Figure 8d shows a quasi periodic orbit moving around galaxy 1 with initial conditions: $x_0=-0.595, y_0=0, p_{x0}=0$, while the values of all the other parameters are as in Fig. 3b. This orbit is characteristic of 3:4 resonance. In Figure 8e a complicated orbit with initial conditions: $x_0=1.53, y_0=0, p_{x0}=0.767$, moving around galaxy 2 is presented. The values of all other parameters are as in Fig. 3b. In Figure 8f we see a quasi periodic orbit around galaxy 1, with initial conditions: $x_0=-1.22, y_0=0, p_{x0}=0.92$. The values of all other parameters for this figure are as in Fig. 4a. Figure 8g shows a figure eight type orbit circulating around galaxy 1, with initial conditions: $x_0=-0.9, y_0=0, p_{x0}=3.305$, while the values of all the other parameters are as in Fig. 5a. A regular orbit starting at the center of galaxy 1, with initial conditions: $x_0=-0.8235, y_0=0, p_{x0}=0$, is given in Figure 8h. The values of all other parameters for this orbit are as in Fig. 5b. All regular orbits were calculated for a time period of 150 time units.

Figure 9a-d shows four different chaotic orbits of the 2D dynamical system. In Figure 9a we can see a chaotic orbit with initial conditions: $x_0=-1.92, y_0=0, p_{x0}=0$, approaching only host galaxy 1. The time interval of the numerical integration for this orbit is 180 time units. Figure 9b shows a chaotic orbit approaching only host galaxy 2. The initial conditions are: $x_0=1.92, y_0=0, p_{x0}=0$, while the time of the numerical integration is 110 time units. In Figure 9c we observe a chaotic orbit approaching both galaxies. The initial conditions are: $x_0=-1.2, y_0=0, p_{x0}=0$, while the time of the numerical integration is 300 time units. So far, one may conclude that there are two kinds of chaotic orbits, as we have mentioned previously. However, numerical simulations of a large number of chaotic orbits (about 1000) with different initial conditions $\left(x_0,p_{x0}\right)$, show that inevitably all chaotic orbits will approach both galaxies, after a certain time interval. Therefore, the crucial factor is the time interval of the numerical integration. This can be shown better in Figure 9d. This chaotic orbit has the same initial conditions as the orbit shown in Fig. 9a, but the time of the numerical integration this time, is equal to 250 time units. We can see that now the orbit approaches both host galaxies. The values of all other parameters for the orbits shown in Fig. 9a-d are as in Fig. 3a. Of course, this phenomenon is observed only in the case when the $(x,p_x)$ phase planes of the two galaxies are connected (see Figs. 3a and 3b). When the centers of the two galaxies located at distances $R$ such as they do not communicate directly in the $(x,p_x)$ phase plane (see Figs. 4a-b and 5a-b) then obviously, there are chaotic orbits approaching only one of the two galaxies, regardless of the time interval of numerical integration.
\begin{figure}[!tH]
\centering
\resizebox{0.5\hsize}{!}{\rotatebox{0}{\includegraphics*{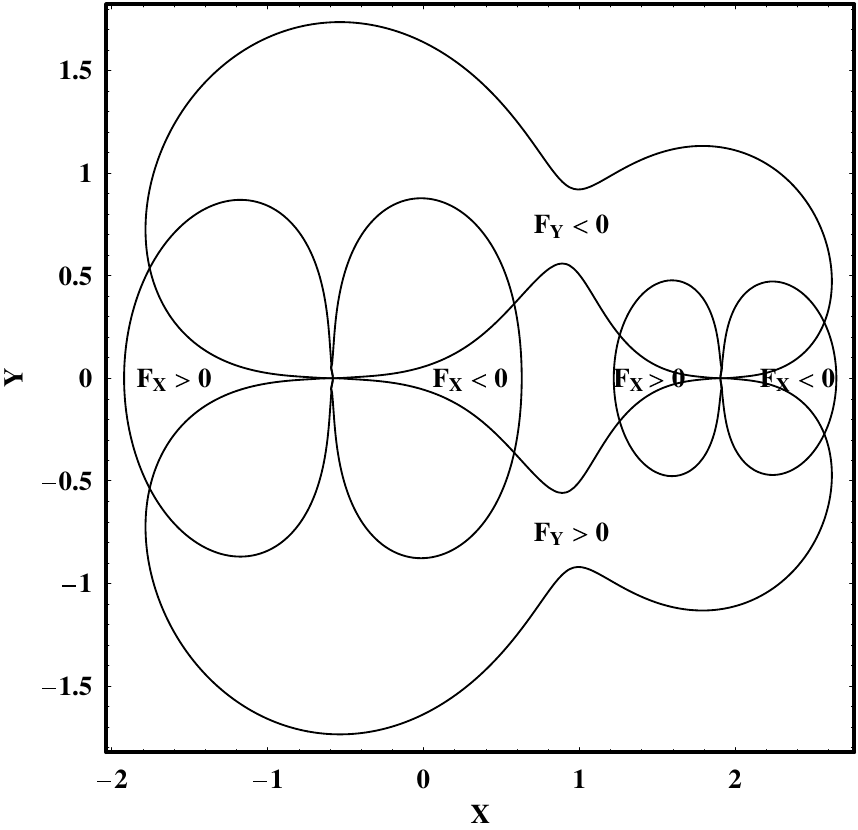}}}
\caption{Contours of the $F_x=const.$ (elliptic shaped) together with the contours $F_y=const.$ (figure eight shaped). Details are given in the text.}
\end{figure}

In what follows we shall come to present some semi-theoretical results, in order to give a more detailed picture of the structure of 2D dynamical system and its behavior. The forces acting on a test particle along the $x$ and $y$ axis are given by the equations
\begin{eqnarray}
F_x &=& -\frac{M_{n1}\left(x-x_1\right)}{\left(r_{a1}^2+c_{n1}^2\right)^{3/2}} -
\frac{M_{d1}\left(x-x_1\right)}{\left[b_1^2+r_{a1}^2+\left(a_1+h_1\right)^2\right]^{3/2}}
- \frac{M_{n2}\left(x-x_2\right)}{\left(r_{a2}^2+c_{n2}^2\right)^{3/2}} \nonumber\\
&-& \frac{M_{d2}\left(x-x_2\right)}{\left[b_2^2+r_{a2}^2+\left(a_2+h_2\right)^2\right]^{3/2}}
+ \Omega_p^2 x - 2\Omega_p \dot{y} \ \ \ , \nonumber\\
F_y &=& -\frac{M_{n1}y}{\left(r_{a1}^2+c_{n1}^2\right)^{3/2}} -
\frac{M_{d1}y}{\left[b_1^2+r_{a1}^2+\left(a_1+h_1\right)^2\right]^{3/2}}
- \frac{M_{n2}y}{\left(r_{a2}^2+c_{n2}^2\right)^{3/2}} \nonumber\\
&-& \frac{M_{d2}y}{\left[b_2^2+r_{a2}^2+\left(a_2+h_2\right)^2\right]^{3/2}}
+ \Omega_p^2 y + 2\Omega_p \dot{x} \ \ \ .
\end{eqnarray}
It is obvious from equations (13) that the strength of both forces increases as the masses of the nuclei or the disks increase or their scale lengths decrease. Figure 10 shows the contours of $F_x=const.$ together with the contours $F_y=const.$. The values of all other parameters are as in Fig. 2. The contours of $F_x=const.$ look like ellipses, while the contours of $F_y=const.$ have a figure eight shape. Inside the first and third ``ellipse" from left to right we have $F_x > 0$, while inside the second and the fourth ``ellipse" we have $F_x < 0$. On the other hand, inside the lower figure eight curve we have $F_y > 0$, while inside the upper figure eight curve we have $F_y < 0$. Looking carefully near each galactic center, we observe that there are areas where $F_x$ is positive and $F_y$ is negative at the same time and vise versa. There are also areas, where both forces are positive or negative at the same time. Therefore, we conclude that near each galactic center, there are strong attractive and repulsive forces acting on the test particle (star). These forces are responsible for the chaotic scattering of the star near each galactic center, leading to chaotic orbits.

\section{STRUCTURE OF THE 3D HAMILTONIAN SYSTEM}

In this Section we shall investigate the regular or chaotic nature of motion in the 3D Hamiltonian system described by equation (10). In order to keep things simple, we shall use our experience gained from the study of the 2D dynamical system, in order to obtain a clear picture regarding the properties of motion in the 3D dynamical model. We are particularly interested to locate the initial conditions in the 3D dynamical system, producing regular or chaotic orbits. A convenient way to obtain this, is to start from $(x,p_x)$ phase planes of the 2D system with the same value of the Jacobi integral, used in the 2D system and described in the previous section. Specifically, the regular or chaotic nature of the 3D orbits is found as follows: we choose initial conditions $\left(x_0,p_{x0},z_0\right), y_0=p_{z0}=0$, such as $\left(x_0,p_{x0}\right)$ is a point on the phase planes of the 2D system. The points $\left(x_0,p_{x0}\right)$ lie inside the limiting curve
\begin{equation}
\frac{1}{2}p_x^2 + \Phi_t(x) = E_{J2} \ \ \ ,
\end{equation}
which is the limiting curve containing all the invariant curves of the 2D system. Thus we take $E_J=E_{J2}$. For this purpose a large number of orbits (about 1000) was computed with initial conditions $\left(x_0,p_{x0},z_0\right)$, where $\left(x_0,p_{x0}\right)$ is a point in the chaotic regions of the $(x,p_x)$ phase planes of Figs. 3a-b, 4a-b and 5a-b, with all permissible values of $z_0$ and $p_{z0}=0$. Remember that, as we are on the phase plane, we have $y_0=0$, while in all cases the value of $p_{y0}$ was obtained from the Jacobi integral (10). All tested orbits were found to be chaotic. Therefore, one concluded that orbits with the initial conditions $\left(x_0,p_{x0}\right)$ be a point in the chaotic regions of the 2D phase planes and for all permissible values of $z_0$, remain chaotic in the 3D system as well.
\begin{figure}[!tH]
\centering
\resizebox{0.8\hsize}{!}{\rotatebox{0}{\includegraphics*{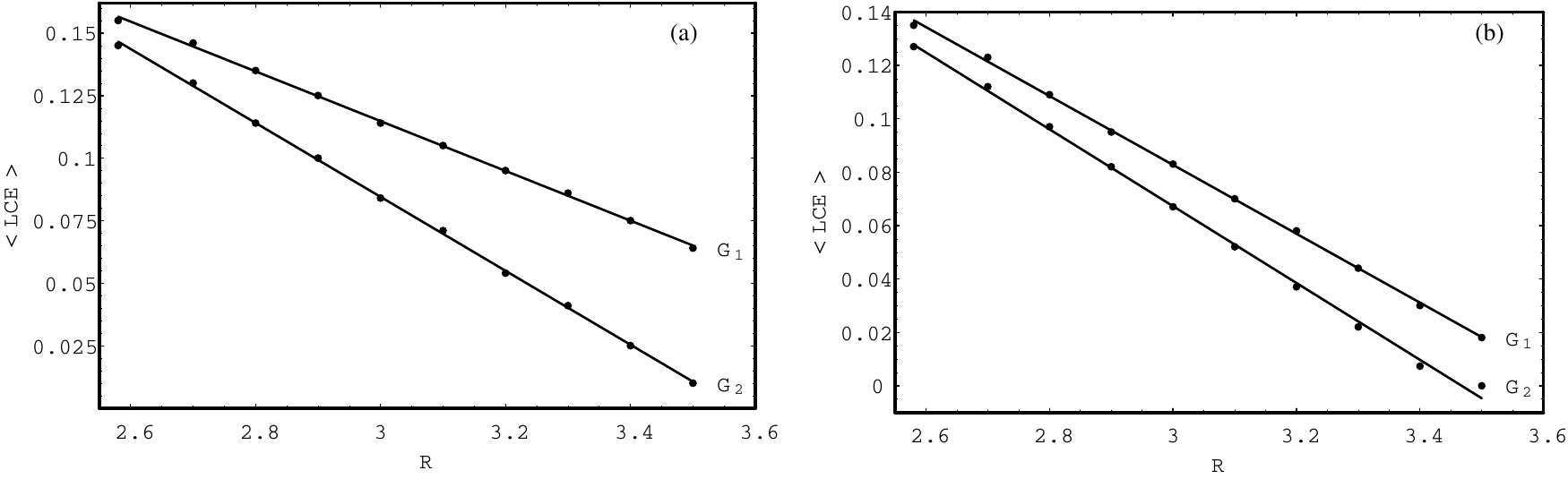}}}
\caption{(a-b): Similar to Fig. 7a-b but for the 3D dynamical system. The values of all the other parameters are given in the text.}
\end{figure}
\begin{figure}[!tH]
\centering
\resizebox{0.9\hsize}{!}{\rotatebox{0}{\includegraphics*{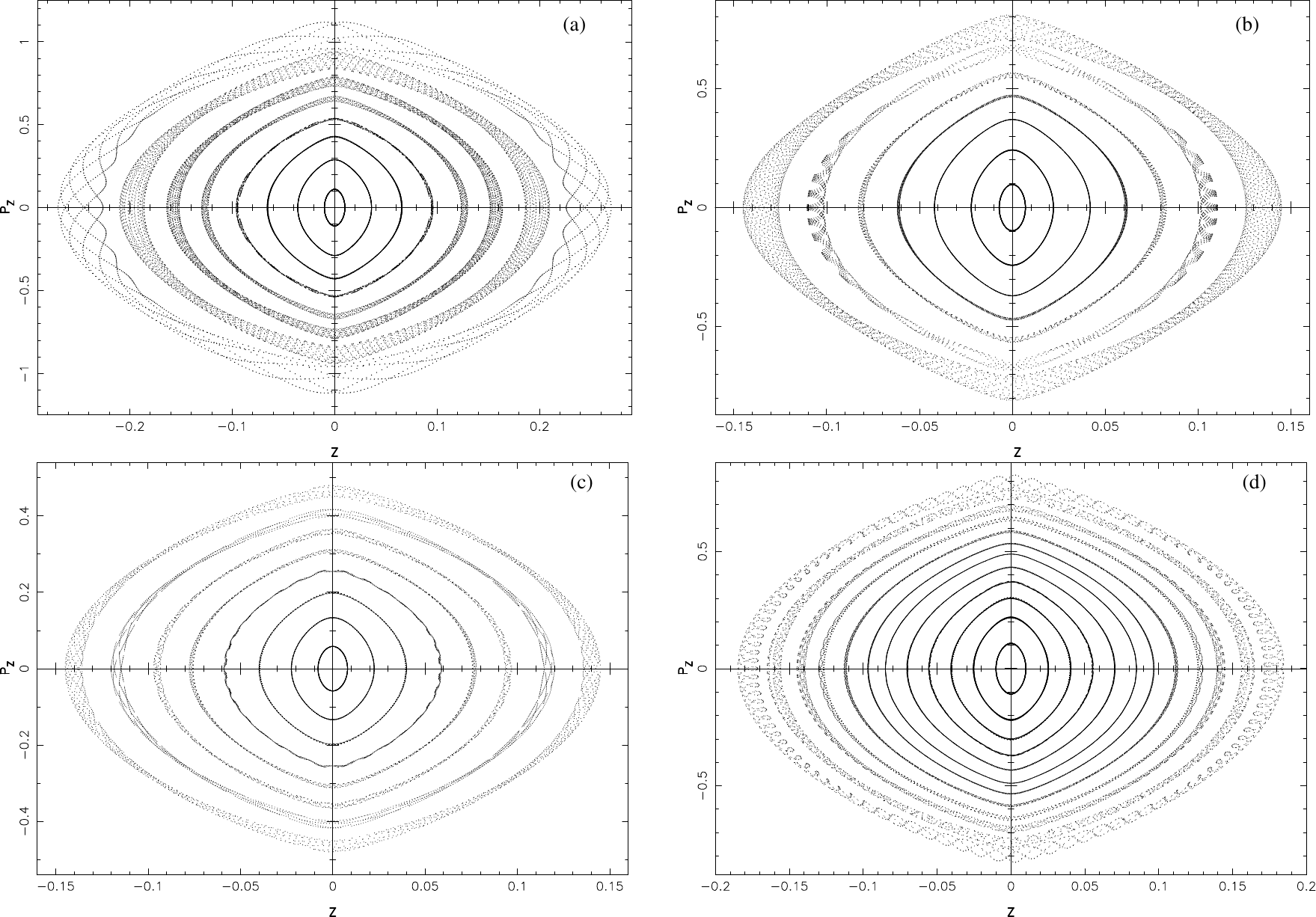}}}
\caption{(a-d): Projections of the sections of 3D orbits with the plane $y=0$ when $p_y>0$. The set of the four dimensional points $\left(x,p_x,z,p_z\right)$ is projected on the $(z,p_z)$ plane.}
\end{figure}

Figure 11a shows a plot of the average value of the maximum Lyapunov Characteristic Exponent - LCE of the 3D system, as a function of the distance between the centers of the two host galaxies. The values of the parameters are: $M_{n1}=0.08, M_{d1}=2.0, M_{n2}=0.04, M_{d2}=0.6$, while the value of the Jacobi integral is $E_J=-2.0$. In this case, the quasars are present in the cores of two host galaxies G1 and G2. The value of the angular frequency $\Omega_p$ is calculated for each particular value of the distance $R$ from equation (3). One can observe, in Fig. 11a, that $< LCE >$ decreases linearly as the distance $R$ increases for both cases (G1 and G2). Figure 11b is similar to Fig. 11a but for the case when the quasars are not present in the cores of the two galaxies. The values of the parameters are: $M_{n1}=0, M_{d1}=2.08, M_{n2}=0, M_{d2}=0.64$, while the value of the Jacobi integral is $E_J=-2.0$. Here again $< LCE >$ decreases linearly, as the distance $R$ increases, for both quiet galaxies G1 and G2. Once more, as we have pointed out in Figs. 7a and 7b regarding the results of the 2D system, for each particular value of the distance $R$, the average value $< LCE >$ of the 3D system, is always smaller when the quasars are not present, in both galaxies G1 and G2. The method we use in order to compute the $< LCE >$ in each case, is the same as described in Figs. 7a-b. For all tested chaotic orbits in the 3D system the initial value of $z_0$ is common and equal to 0.1. Here we must point out, that if we compare the plots of the 3D system shown in Figs. 11a-b with those of the 2D system shown in Figs. 7a-b, we observe that in each case (active and quiet galaxies respectively) the average values of the LCEs of the 3D system are smaller than the corresponding of the 2D system.

Our next step, is to study the character of orbits with initial conditions $\left(x_0,p_{x0},z_0\right), y_0=p_{z0}=0$, such as $\left(x_0,p_{x0}\right)$ is a point in the regular regions of Figs. 3a-b, 4a-b and 5a-b. The phase space of a conservative system of three degrees of freedom has six dimensions, i.e. in Cartesian coordinates $\left(x, y, z, \dot{x}, \dot{y}, \dot{z}\right)$. For a given value of the Jacobi integral a trajectory lies on a 5-dimensional manifold. In this manifold the surface of section is 4-dimensional. This does not allow us to visualize and interpret directly the structure and the properties of the phase space in dynamical systems of three degrees of freedom. One way to overcome this problem, is to project the surface of section to space with lower dimensions. In fact we will apply the method introduced by Pfenniger (1984) (see also Revaz and Pfenniger, 2001). We take sections in the plane $y=0, p_y >0$ of 3D orbits, whose initial conditions differ from the plane parent periodic orbits only by the $z$ component. The set of the resulting 4-dimensional points in $\left(x,p_x,z,p_z\right)$ phase space are projected on the $\left(z,p_z\right)$ plane. If the projected points lie on a well defined curve, lets call it an ``invariant curve", then the motion is regular, while if not, the motion is chaotic. The projected points on the $\left(z,p_z\right)$ plane shows nearly invariant curves around the periodic points at $z=0, p_z=0$, as long as the coupling is weak. When the coupling is stronger the corresponding projections in $\left(z,p_z\right)$ plane shows increasing departure of the plane periodic point, up to making a direct orbit a retrograde one and vise versa. Here we must define what one means by direct and retrograde 3D orbit. If consequents in the $\left(z,p_z\right)$ section of the 3D orbit drop in one of the two domains of the corresponding section of 2D orbits at the same value of the Jacobi integral $E_J$, we can distinguish between direct and retrograde motion. Orbits which visit both domains are intermittently direct or retrograde.
\begin{figure}[!tH]
\centering
\resizebox{0.5\hsize}{!}{\rotatebox{0}{\includegraphics*{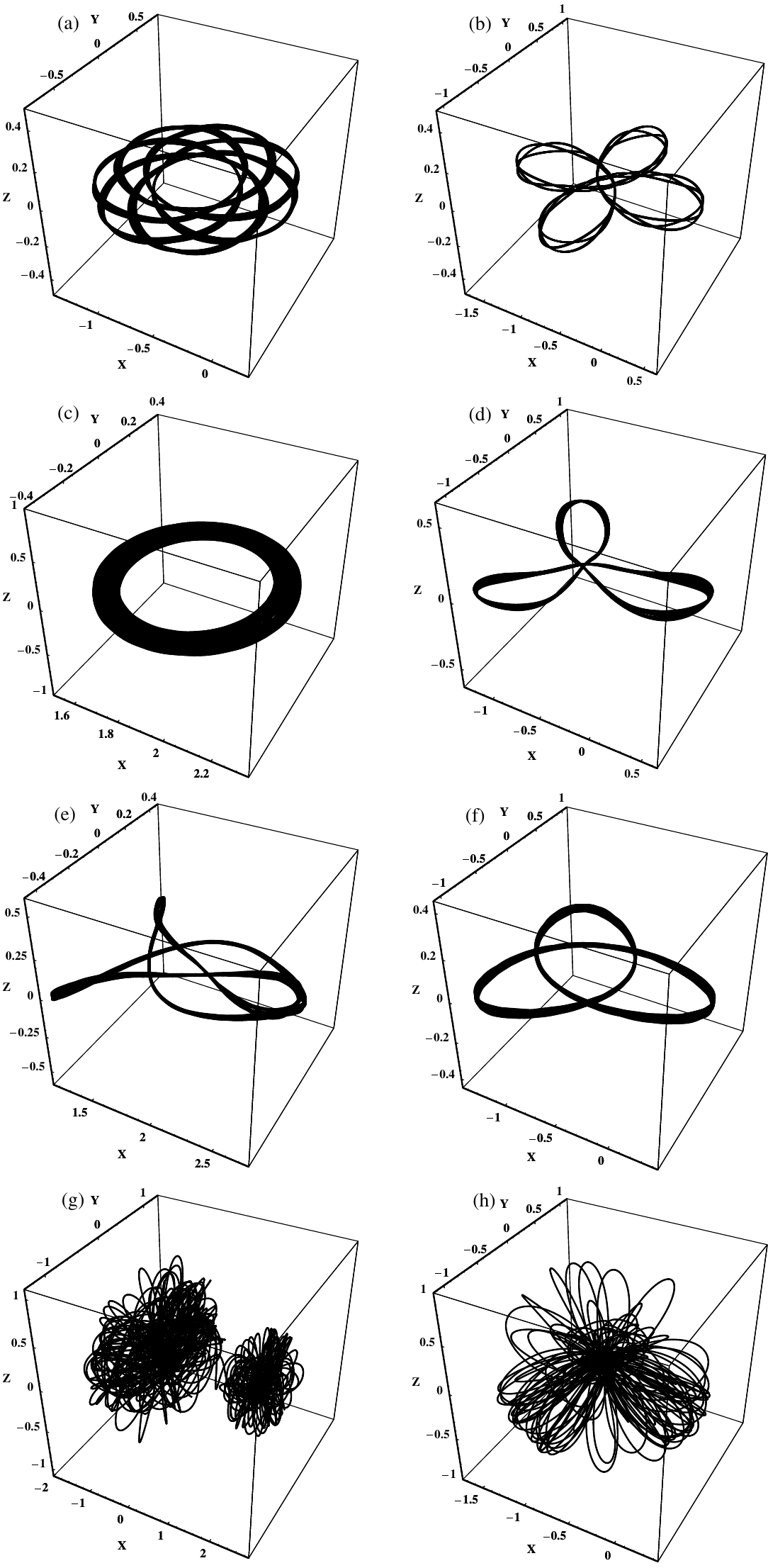}}}
\caption{(a-h): Eight orbits of the 3D dynamical system. The initial conditions and more details regarding the values of all the other parameters are given in the text.}
\end{figure}

Figure 12a-d shows four typical $\left(z,p_z\right)$ sections of 3D orbits, starting with initial conditions close to different periodic points on the $\left(x,p_x\right)$ phase planes of the 2D system. In order to obtain the results shown in Fig. 12a we have taken the point $\left(x_0,p_{x0}\right)=(-0.05,0)$ representing approximately the position of the retrograde periodic point in the $\left(x,p_x\right)$ phase plane of Fig. 3a for the active host galaxy 1. Similarly in Fig. 12b we can observe the $\left(z,p_z\right)$ projections near the retrograde periodic point of quiet galaxy 2, shown in Fig. 3b. The position of the periodic point is $\left(x_0,p_{x0}\right)=(2.25,0)$. Moreover in Fig. 12c we see the $\left(z,p_z\right)$ projections near the direct periodic point of active host galaxy 1, shown in Fig. 4a. The position of the periodic point is $\left(x_0,p_{x0}\right)=(-1.48,0)$. One more example is given in Fig. 12d, where one can see the $\left(z,p_z\right)$ projections near the retrograde periodic point of active host galaxy 1, shown in Fig. 5a. The position of the periodic point this time is $\left(x_0,p_{x0}\right)=(-0.28,0)$. Note that in all cases the numerical results indicate, that for small values of $z_0$ the motion is regular, while for larger values of $z_0$ the motion become chaotic. Numerical calculations not given here, suggest that the above method can be applied in all regular regions around each retrograde or direct stable periodic point of Figs. 3a-b, 4a-b and 5a-b.

We must emphasize that the results shown in Fig. 12a-d are rather qualitative and can be considered as an indication that the transition from regularity to chaos in 3D orbits, occurs as the value of $z_0$ increases. In order to form a more complete and accurate view of the phase space in the 3D system, we computed a large number of 3D orbits (approximately 1000) near in each periodic point of the $\left(x,p_x\right)$ phase planes of 2D system, for different initial conditions $\left(x_0,p_{x0}\right)$ and also for different values of $z_0$. Our target was to determine the average minimum value of $z_0$ for which the nature of a 3D orbit changes from regular to chaotic. Table 1 shows the value $< z_{min} >$ near the direct and retrograde stable periodic points of Figs. 3a-b, 4a-b and 5a-b, for three different values of distance between the centers of the two host galaxies $R$. From Table 1 we can induce three important results. (i). In every case (Active or Quiet galaxy) the minimum value of $z_0$ in the regions near the retrograde stable periodic points, is always larger than the corresponding of the regions of the direct periodic points. (ii). If we compare each set of galaxies regarding their nuclear activity, (Active G1,2 with Quiet 1,2) for the same distance $R$, we observe that when the quasars are not present at the galactic cores (Quiet - non Active galaxies), the 3D orbits can approach higher values of $z_0$ and remain regular. On the other hand, when the quasars are present, the value of $< z_{min} >$ is smaller. (iii). As the distance between the centers of the galaxies increases, the minimum initial value of $z_0$ for which a 3D can remain regular, increases in both cases (Active and Quiet galaxies). This means that as the two galaxies are in large distances, their mutual interactions are weak enough and therefore the majority of 3D orbits are ordered. Moreover, in large distances one can conclude that the main responsible for the observed chaotic motion is the nuclear activity of the quasars in the cores of each host galaxy.
\begin{table}[ht]
\caption{Average value of minimum $z_0$ near the direct and retrograde periodic points, for different values of the distance $R$.}
\centering
\begin{tabular}{|c|c|c||c|}
\hline
\textbf{Distance} & \textbf{Case} & \textbf{Region} & $< z_{min} >$ \bigstrut[t] \\
\hline \hline
\multirow{8}[8]{*}{$R=2.5$} &
\multirow{2}[4]{*}{Active G1}
          & Direct
                       & - \bigstrut[t] \\ \cline{3-4}
&         & Retrograde & 0.125 \bigstrut[t] \\ \cline{2-4} &
\multirow{2}[4]{*}{Active G2}
          & Direct
                       & 0.042 \bigstrut[t] \\ \cline{3-4}
&         & Retrograde & 0.117 \bigstrut[t] \\ \cline{2-4} &
\multirow{2}[4]{*}{Quiet G1}
          & Direct
                       & - \bigstrut[t] \\ \cline{3-4}
&         & Retrograde & 0.128 \bigstrut[t] \\ \cline{2-4} &
\multirow{2}[4]{*}{Quiet G2}
          & Direct
                       & 0.047 \bigstrut[t] \\ \cline{3-4}
&         & Retrograde & 0.086 \bigstrut[t] \\
\hline
\multirow{8}[8]{*}{$R=3.0$} &
\multirow{2}[4]{*}{Active G1}
          & Direct
                       & 0.092 \bigstrut[t] \\ \cline{3-4}
&         & Retrograde & 0.131 \bigstrut[t] \\ \cline{2-4} &
\multirow{2}[4]{*}{Active G2}
          & Direct
                       & 0.105 \bigstrut[t] \\ \cline{3-4}
&         & Retrograde & 0.121 \bigstrut[t] \\ \cline{2-4} &
\multirow{2}[4]{*}{Quiet G1}
          & Direct
                       & 0.098 \bigstrut[t] \\ \cline{3-4}
&         & Retrograde & 0.136 \bigstrut[t] \\ \cline{2-4} &
\multirow{2}[4]{*}{Quiet G2}
          & Direct
                       & 0.109 \bigstrut[t] \\ \cline{3-4}
&         & Retrograde & 0.124 \bigstrut[t] \\
\hline
\multirow{8}[8]{*}{$R=3.5$} &
\multirow{2}[4]{*}{Active G1}
          & Direct
                       & 0.127 \bigstrut[t] \\ \cline{3-4}
&         & Retrograde & 0.138 \bigstrut[t] \\ \cline{2-4} &
\multirow{2}[4]{*}{Active G2}
          & Direct
                       & 0.122 \bigstrut[t] \\ \cline{3-4}
&         & Retrograde & 0.126 \bigstrut[t] \\ \cline{2-4} &
\multirow{2}[4]{*}{Quiet G1}
          & Direct
                       & 0.132 \bigstrut[t] \\ \cline{3-4}
&         & Retrograde & 0.141 \bigstrut[t] \\ \cline{2-4} &
\multirow{2}[4]{*}{Quiet G2}
          & Direct
                       & 0.129 \bigstrut[t] \\ \cline{3-4}
&         & Retrograde & 0.137 \bigstrut[t] \\
\hline
\end{tabular}
\end{table}

So far we have seen that 3D orbits with initial conditions $\left(x_0,p_{x0},z_0\right)$, such as $\left(x_0,p_{x0}\right)$ is a point in the chaotic regions of the 2D system, for all permissible values of $z_0$ are chaotic. On the other hand, the nature (ordered or chaotic) of 3D orbits with initial conditions $\left(x_0,p_{x0},z_0\right)$, such as $\left(x_0,p_{x0}\right)$ is a point in the regular regions around the stable direct or retrograde periodic points of the 2D system, depends on the particular value of $z_0$, as we can see in Table 1. We did not feel that it was necessary to try to define the values of $< z_{min} >$ for each regular region of the 2D system corresponding to secondary resonances which are represented by multiple small islands in the $(x,p_x)$ phase planes, shown in Figs. 3a-b, 4a-b and 5a-b. Numerical results indicate that 3D orbits with initial conditions $\left(x_0,p_{x0},z_0\right)$, such as $\left(x_0,p_{x0}\right)$ is a point in the regular regions corresponding to secondary resonances of the 2D system, remain regular for $< z_{min} > \simeq 0.047$, while for larger values of $z_0$, they change their character from regular to chaotic.

Figure 13a-h depicts eight orbits of the 3D dynamical system. Fig. 13a shows a regular quasi periodic orbit circulating around active galaxy 1, with initial conditions: $x_0=0.2, y_0=0, z_0=0.01, p_{x0}=p_{z0}=0$, while the value of $p_{y0}$ is always found from the Jacobi integral (10). In Fig. 13b we observe a 3D quasi periodic orbit characteristic of the 3:3 resonance. This orbit has initial conditions: $x_0=0.52, y_0=0, z_0=0.03, p_{x0}=p_{z0}=0$ and going around active galaxy 1. Fig. 13c shows a 3D orbit with initial conditions: $x_0=2.2, y_0=0, z_0=0.03, p_{x0}=p_{z0}=0$, circulating around active galaxy 2. In Fig. 13d we see quasi periodic orbit, with initial conditions: $x_0=-0.595, y_0=0, z_0=0.015, p_{x0}=p_{z0}=0$, moving around quiet galaxy 1. A complicated 3D resonant orbit with initial conditions: $x_0=1.53, y_0=0, z_0=0.01, p_{x0}=0.767, p_{z0}=0$, moving around quiet galaxy 2, is shown in Fig. 13e. Fig. 13f shows a quasi periodic orbit circulating around active galaxy 1. This orbit has initial conditions: $x_0=-1.22, y_0=0, z_0=0.01, p_{x0}=0.92, p_{z0}=0$. In Fig. 13g we observe a 3D chaotic orbit with initial conditions: $x_0=-1.2, y_0=0, z_0=0.1, p_{x0}=p_{z0}=0$, approaching arbitrarily both active galaxies. It is interesting to note that near the more massive host galaxy (G1) the orbit is deflected to higher values of $z$, while near the less massive host galaxy (G2), the orbit stays close to the disk. Fig. 13h shows an orbit with the same initial conditions as in Fig. 13a but for $z_0=0.2$. The orbit has now become chaotic and goes arbitrarily close to active host galaxy 1. This orbit shows, from another point of view, that 3D orbits starting near the stable periodic points (direct or retrograde) of the 2D system remain regular only for small values of $z_0$. We must note that in all 3D orbits shown in Fig. 13a-g the initial conditions $\left(x_0,p_{x0}\right)$ and the values of all the other parameters are as in the corresponding 2D orbits shown in Figs. 8a-f and 9c. Moreover, we observe that all regular 3D orbits shown in Fig. 13a-f, stay neat to the galactic plane and therefore support the disk of each host galaxy. The numerical integration time for all regular 3D orbits shown in Fig. 13a-f, is 200 time units, while for the chaotic 3D orbit shown in Fig. 13g is 500 time units. We should also mention that the phenomenon regarding the time integration and its influence on the structure of an orbit, which was discussed in detail in the 2D system of the previous section, was also observed in the 3D system. For instance the chaotic orbit shown in Fig. 13h was integrated for a time interval of 220 time units and it moves only around host active galaxy 1. But if we integrate numerically this orbit for a longer time interval, we will observe that the orbit will gradually form a shape like the one shown in Fig. 13g, which means that it will approaches both host galaxies.
\begin{figure}[!tH]
\centering
\resizebox{0.8\hsize}{!}{\rotatebox{0}{\includegraphics*{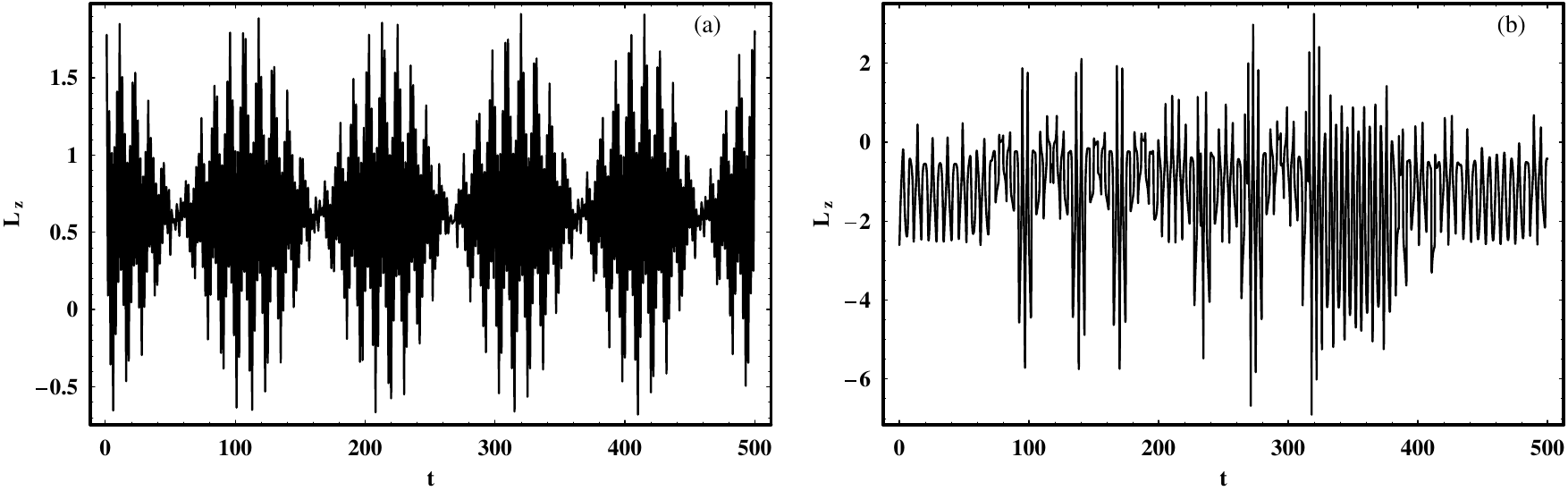}}}
\caption{(a-b): A plot of $L_z$ component of the total angular momentum vs. time for (a-left): a regular 3D orbit and (b-right): a chaotic 3D orbit.}
\end{figure}
\begin{figure}[!tH]
\centering
\resizebox{0.8\hsize}{!}{\rotatebox{0}{\includegraphics*{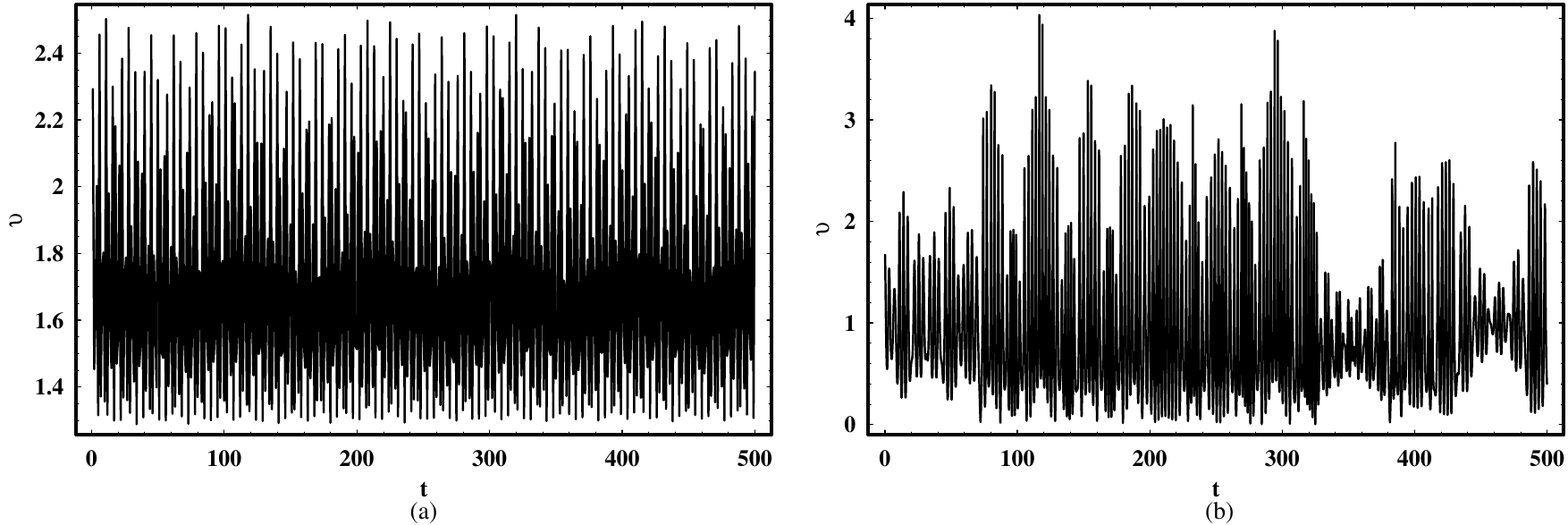}}}
\caption{(a-b): (a-left): The total velocity profile of the 3D orbit shown in Fig. 13a. We observe a nearly periodic pattern and (b-right): The total velocity profile for the chaotic 3D orbit shown in Fig. 13g. In this case there are abrupt changes in the profile's pattern indicating chaotic motion.}
\end{figure}
\begin{figure}[!tH]
\centering
\resizebox{0.75\hsize}{!}{\rotatebox{0}{\includegraphics*{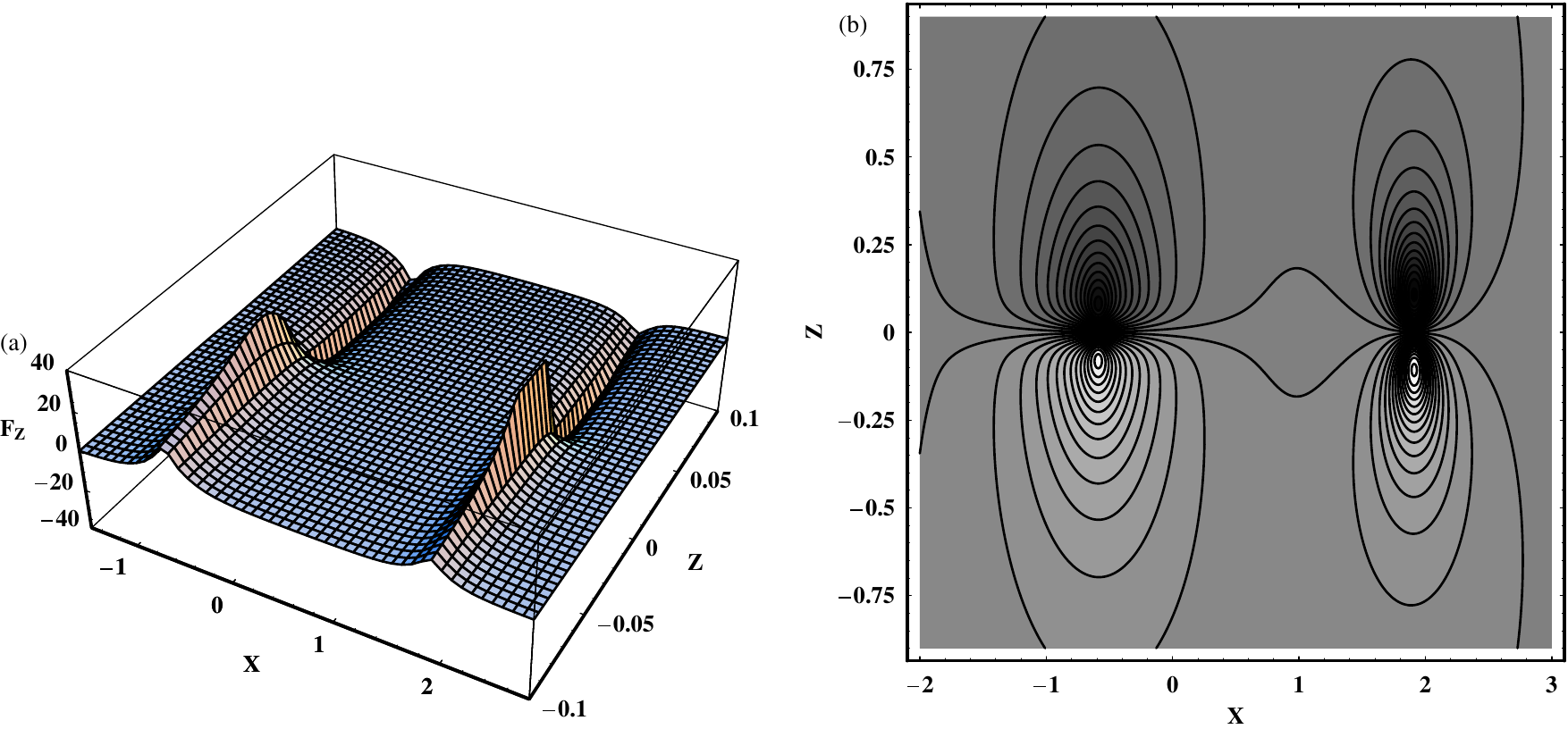}}}
\caption{(a-b): (a-left): A 3D plot of the value of the $F_z$ force on the $(x,z)$ plane and (b-right): Contours of the projections $F_z=const.$ on the $(x,z)$ plane. Lighter colors indicate higher values of $F_z$. For positive values of $z$ the $F_z$ force is negative, while for negative values of $z$ the $F_z$ force is positive.}
\end{figure}
\begin{figure}[!tH]
\centering
\resizebox{0.5\hsize}{!}{\rotatebox{0}{\includegraphics*{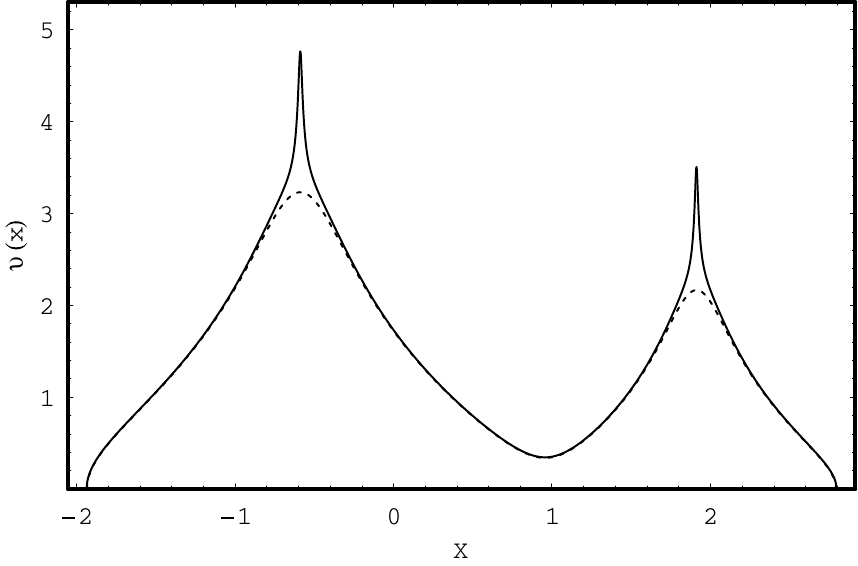}}}
\caption{A plot of the total velocity $\upsilon (x)$ vs. distance $x$ using Eq. (18) for the two cases (Active and Quiet galaxies).}
\end{figure}

The physical parameter playing an important role on the orbital behavior of the stars, is the $L_z$ component of the total angular momentum. From our previous experience, we know that low angular momentum stars on approaching a dense, massive nucleus are scattered off the galactic plane, displaying chaotic motion (Caranicolas and Innanen, 1991; Caranicolas and Papadopoulos, 2003, Caranicolas and Zotos, 2010; Zotos, 2011). Of course here in 3D phase space things are more complicated than in axially symmetric dynamical models, where the $L_z$ component was conserved. As the motion takes place in a rotating non axially symmetric system, the $L_z$ component, which is given by
\begin{equation}
L_z = x \dot{y} - \dot{x} y - \Omega_p \left(x^2 + y^2\right) \ \ \ ,
\end{equation}
is not conserved. Nevertheless, we can compute numerically the mean value $< L_z >$ of $L_z$ using the formula
\begin{equation}
< L_z > = \frac{1}{n} \displaystyle\sum\limits_{i=0}^n L_{zi} \ \ \ .
\end{equation}
Our numerical calculations suggest that the chaotic orbits have low values of $< L_z >$, while regular orbits obtain high values of $< L_z >$. Figure 14a shows a plot of the evolution of $L_z$ component with the time for the regular orbit of Fig. 13a. In this case we observe that $L_z$ is nearly a periodic function of time, while $< L_z > = 0.6274$. Figure 14b is similar to 14a but the chaotic orbit shown in Fig. 13g. Here one can see abrupt changes of $L_z$ during the chaotic motion, while for this chaotic orbit we have $< L_z > = -1.8266$. In both cases the time interval of the numerical integration is 500 time units, while $n =10^4$.

Moreover, it would be of particular interest to study the structure of the velocity profile, that is the plot of the total velocity of a test particle (star), $\upsilon=\sqrt{\dot{x}^2 + \dot{y}^2 + \dot{z}^2}$ as a function of time, for ordered and chaotic motion. Figure 15a shows the velocity profile for a time period of 500 time units corresponding to the regular 3D orbit shown in Fig. 13a. Here the velocity profile is quasi periodic and the maximum value of the velocity is about 490 km/s. This suggest that regular 3D motion is made in small velocities. Figure 15b shows the velocity profile for a time period of 500 time units corresponding to the chaotic 3D orbit shown in Fig. 13g. In this case we can discuss two aspects. First, the velocity obtains high values up to 790 km/s and second the velocity profile appears to be highly asymmetric displaying abrupt changes and large deviations between the maxima and also between the minima in the $\left[\upsilon - t \right]$ plot. Therefore, we can conclude that the chaotic 3D motion is made in high and abruptly changing velocities.

At this point, we shall present some semi-theoretical results, in order to give a more detailed and complete picture of the structure of 3D dynamical system and its behavior. The method is similar to that we have used in the 2D system of the previous section. The force acting on a test particle along the $z$ axis is given by the equation
\begin{equation}
F_z = -\frac{M_{n1}z}{\left(r_1^2+c_{n1}^2\right)^{3/2}} -
\frac{M_{d1}\left(a_1+r_{z1}\right)z}{r_{z1}\left[b_1^2+r_{a1}^2+\left(a_1+r_{z1}\right)^2\right]^{3/2}}
- \frac{M_{n2}z}{\left(r_2^2+c_{n2}^2\right)^{3/2}} -
\frac{M_{d2}\left(a_2+r_{z2}\right)z}{r_{z2}\left[b_2^2+r_{a2}^2+\left(a_2+r_{z2}\right)^2\right]^{3/2}} \ \ \ ,
\end{equation}
where $r_{z1}=\sqrt{h_1^2+z^2}$ and $r_{z2}=\sqrt{h_2^2+z^2}$. It is obvious from equation (17) that the strength of the $F_z$ force increases as the masses of the nuclei or the disks increase or their scale lengths decrease. Figure 16a shows a 3D plot of the value of the $F_z$ force on the $(x,z)$ plane. Figure 16b depicts the contours of the projections $F_z = const.$ on the $(x,z)$ plane. We observe that for positive values of $z$ the $F_z$ force is negative, while for negative values of $z$ the value of the $F_z$ is positive, near both galactic cores. Therefore, lighter colors on the lower half part of the $(x,z)$ plane indicate higher values of the $F_z$ force, while darker colors on the upper half part of the same plane indicate lower values of the $F_z$ force. One can observe that near each active galactic core which host a quasar, the test particle experiences a very strong $F_z$ force. The values of all the parameters in Fig. 16a-b, are as in Fig. 2.

Last but not least, it would be also interesting to investigate the behavior of the velocity near each galactic core, as a function of the distance $x$. In order to do that, we consider the limiting curve, that is, the curve containing all the invariant curves on the $(x,p_x)$ phase plane. This can be obtained, if we set $y=z=p_y=p_z=0$ in equation (10) yielding to
\begin{equation}
\frac{1}{2}p_x^2 = \frac{1}{2}\upsilon^2 = E_J - \Phi_t (x) \ \ \ ,
\end{equation}
where we have set $p_x=\upsilon$ because at the limiting curve the $p_x$ velocity is the total velocity (see also Papadopoulos and Caranicolas, 2006). In Figure 17 we can observe two plots of the velocity $\upsilon$ as a function of the $x$, derived using Eq. (18). The values of all the parameters are as in Fig. 2. The dashed line corresponds to the case of quiet galaxies, while the solid line corresponds to the case when the galaxies host the quasars in their cores and therefore are active. We see that the velocity is about the same for all values of $x$, except near each galactic core, where higher velocities correspond to active galactic centers, hosting massive and dense quasars.

\section{EVOLUTION OF 3D ORBITS IN THE TIME-DEPENDENT MODEL}

Let us now come to follow the evolution of 3D orbits, as mass is transported from the disks of the quiet galaxies to their centers. By this procedure a massive and dense quasar is developed in the central regions of each galaxy. The mass transport is linear following the equations
\begin{eqnarray}
M_{n1f} = M_{n1i} + k_1t, \ \ \ M_{d1f} = M_{d1i} - k_1t \ \ \ , \nonumber \\
M_{n2f} = M_{n2i} + k_2t, \ \ \ M_{d2f} = M_{d2i} - k_2t \ \ \ ,
\end{eqnarray}
where $M_{n1i}=M_{n2i}=0$ and $M_{d1i}=2.08, M_{d2i}=0.64$, are the initial values of the mass of the nuclei and the disks respectively, while $k_1$ and $k_2$ are positive parameters. As in Caranicolas and Innanen (2009), we assume that the linear rate described by relations (19) is slow compared to the orbital period of the binary system and therefore it is adiabatic. This is true because the mass transportation period is 100 time units, while the orbital period is about three orders of magnitude smaller. It is also assumed that the transportation stops when the mass of the quasar in each galactic core takes the value $M_{n1f}=0.08$ or $M_{n2f}=0.04$. It is well known that the shape of 3D orbits, sometimes is inconclusive or misleading. In order to overcome this drawback, we have decided to use a highly accurate method such as the maximum Lyapunov Characteristic Exponent - LCE. The main advantage of this method is that it uses certain and objective numerical thresholds beyond which we can distinguish between ordered and chaotic motion. On the other hand this method is very time consuming, as it needs time intervals of numerical integration of order of $10^5$ time units, in order to give reliable and definitive results regarding the nature of a 3D orbit.
But this is a ``price" we can afford to pay.
\begin{figure}[!tH]
\centering
\resizebox{0.8\hsize}{!}{\rotatebox{0}{\includegraphics*{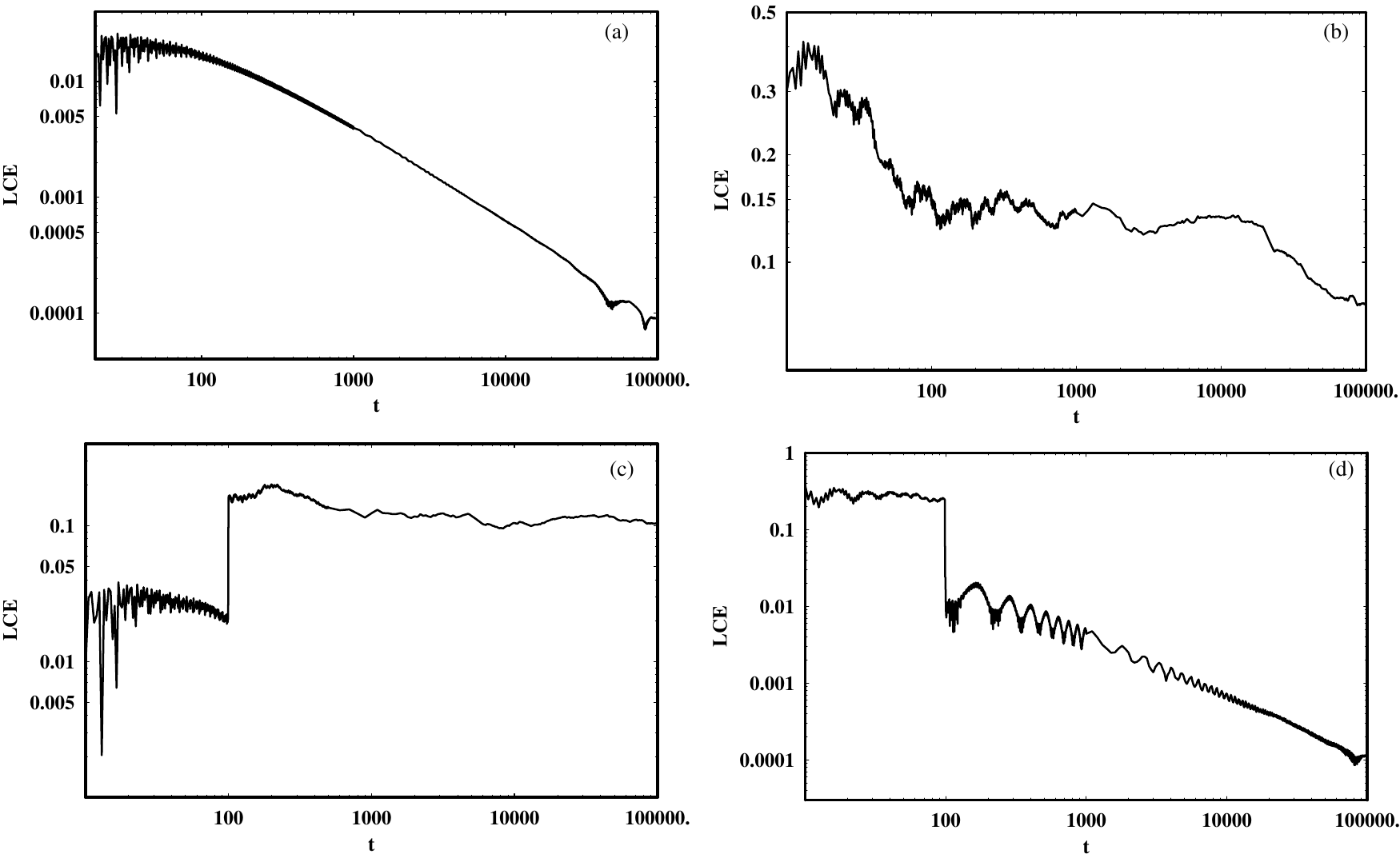}}}
\caption{(a-d): Evolution of the maximum LCEs of four different 3D orbits, following the first set of Eqs. (19). By this procedure a massive quasar is formed in the core of host galaxy 1. (a-upper left): The orbit stars as a regular and remains regular, (b-upper right): the orbits starts as chaotic and remains chaotic, (c-down left): a 3D orbit which stars as regular but after 100 time units when the evolution stops it becomes chaotic and (d-down right): a 3D orbit starting as chaotic but after the evolution it changes its nature to regular. Details are given in the text.}
\end{figure}

Figure 18a-d shows the evolution of four different 3D orbits, as the total mass distribution of the dynamical system changes with time, following the first set of equations (19), regarding host galaxy 1. For all orbits shown in Fig. 18a-d, the initial value of the Hamiltonian (10) is $E_J=-2.0$, while $k_1=0.0008$. In Figure 18a we can see the evolution of the maximum Lyapunov Characteristic Exponent for a 3D orbit and for a time period of $10^5$ time units, as the galaxy 1 evolves following the first set of equations (19). The initial conditions are: $x_0=-0.46, y_0=0, z_0=0.06, p_{x0}=0, p_{z0}=0$, while the value of $p_{y0}$ is found from the Hamiltonian (10) in all cases. The values of all the other parameters are as in Fig. 4b. When $t=100$ time units, the mass of the developed quasar in the core of galaxy 1 is $M_{n1f}=0.08$ and the evolution stops. The value of the Hamiltonian is now $E_J=-2.00032$. The profile of the LCE clearly indicates that this orbit starts as a regular and remains regular, during the galactic evolution. Figure 18b is similar to Fig. 18a. This orbit has initial conditions: $x_0=-0.88, y_0=0, z_0=0.12, p_{x0}=1.25, p_{z0}=0$. The values of all the other parameters are as in Fig. 4b. When $t=100$ time units, the mass of the developed quasar in the core of galaxy 1 is $M_{n1f}=0.08$ and the evolution stops. The value of the Hamiltonian is now $E_J=-2.00044$. In this case, the profile of the LCE clearly indicates that this orbit starts as a chaotic 3D orbit and remains chaotic, during the galactic evolution. In Figure 18c we observe the evolution of the LCE of a 3D orbit with initial conditions: $x_0=-0.64, y_0=0, z_0=0.03, p_{x0}=0, p_{z0}=0$. The values of all the other parameters are as in Fig. 4b. After time interval of 100 time units the mass of the quasar in the core of galaxy 1 is $M_{n1f}=0.08$ and the galactic evolution stops. The Hamiltonian settled to the value $E_J=-2.00027$. The profile of the LCE given in Fig. 18c shows that the orbit stars as a regular 3D orbit but after the galactic evolution it becomes a chaotic one. It is evident that, if mass transport was not present, the orbit would have remain regular. The presence of the quasar in the core of galaxy 1, has changed the character of the 3D orbit, from regular to chaotic. Figure 18d depicts the evolution of the LCE of a 3D orbit with initial conditions: $x_0=-0.93, y_0=0, z_0=0.045, p_{x0}=0, p_{z0}=0$. The values of all the other parameters are as in Fig. 5b. After time interval of 100 time units the mass of the quasar in the core of galaxy 1 is $M_{n1f}=0.08$ and the galactic evolution stops. The Hamiltonian settled to the value $E_J=-2.00015$. The profile of the LCE given in Fig. 18d indicates that this orbit stars as a chaotic 3D orbit but after the galactic evolution it becomes an ordered one. Therefore, it is evident that, if mass transport was not present, the orbit this time would have remain chaotic. In this case, the presence of the quasar in the core of galaxy 1, has changed the nature of the 3D orbit, from chaotic to regular.
\begin{figure}[!tH]
\centering
\resizebox{0.8\hsize}{!}{\rotatebox{0}{\includegraphics*{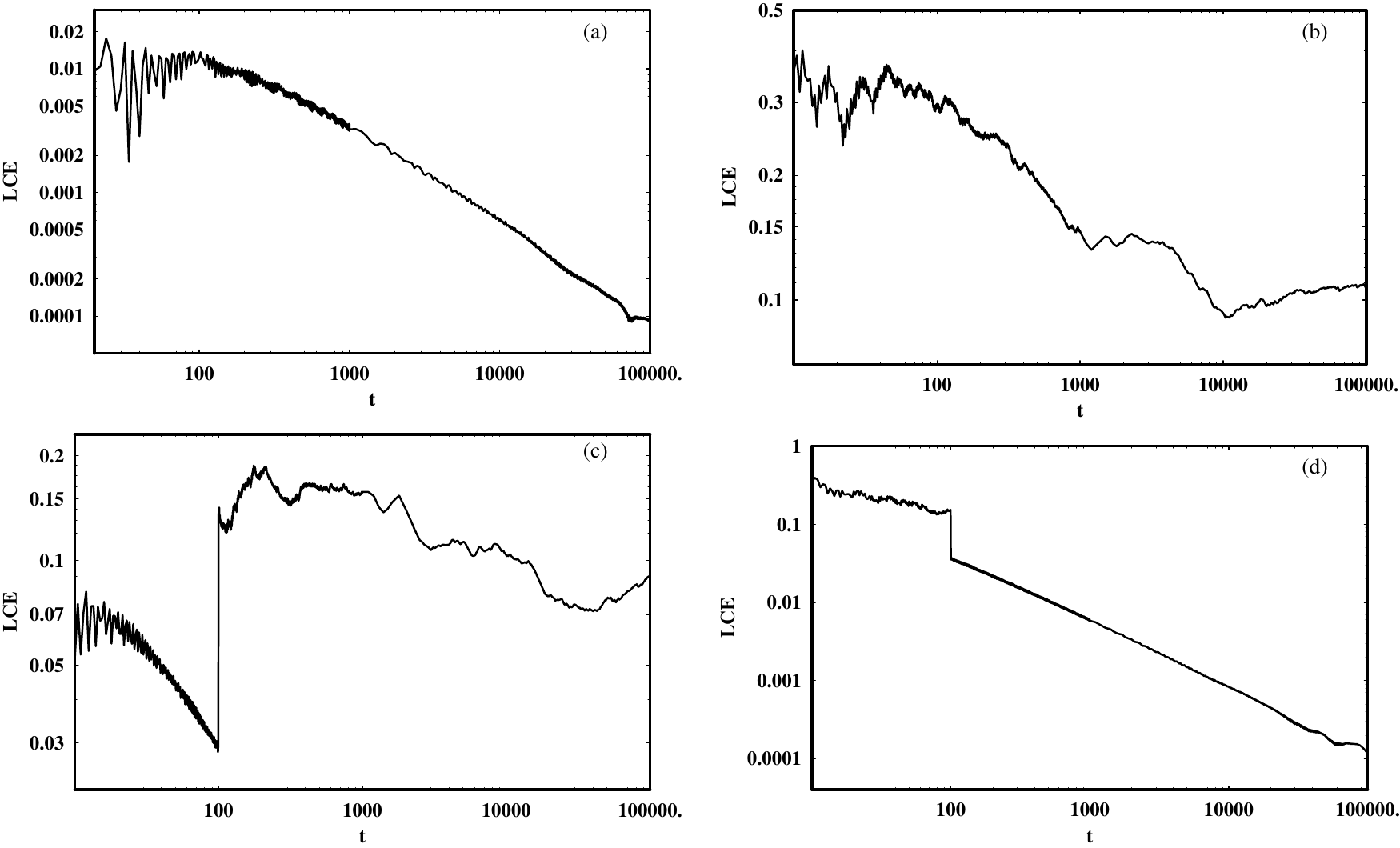}}}
\caption{(a-d): Evolution of the maximum LCEs of four different 3D orbits, following the second set of Eqs. (19). By this procedure a massive quasar is formed in the core of host galaxy 2. (a-upper left): The orbit stars as a regular and remains regular, (b-upper right): the orbits starts as chaotic and remains chaotic, (c-down left): a 3D orbit which stars as regular but after 100 time units when the evolution stops it becomes chaotic and (d-down right): a 3D orbit starting as chaotic but after the evolution it changes its nature to regular. Details are given in the text.}
\end{figure}

In Figure 19a-d one can observes the evolution of four different 3D orbits, as the total mass distribution of the dynamical system changes with time, following the second set of equations (19), regarding host galaxy 2. For all orbits shown in Fig. 19a-d, the initial value of the Hamiltonian (10) is $E_J=-2.0$, while $k_2=0.0004$. In Figure 19a we can see the evolution of the maximum Lyapunov Characteristic Exponent for a 3D orbit and for a time period of $10^5$ time units, as the galaxy 2 evolves following the second set of equations (19). The initial conditions are: $x_0=2.25, y_0=0, z_0=0.035, p_{x0}=0, p_{z0}=0$, while the value of $p_{y0}$ is found from the Hamiltonian (10) in all cases. The values of all the other parameters are as in Fig. 4b. When $t=100$ time units, the mass of the developed quasar in the core of galaxy 2 is $M_{n2f}=0.04$ and the evolution stops. The value of the Hamiltonian is now $E_J=-2.00021$. The profile of the LCE clearly indicates that this orbit starts as a regular and remains regular, during the galactic evolution. Figure 19b is similar to Fig. 19a. This orbit has initial conditions: $x_0=2.05, y_0=0, z_0=0.10, p_{x0}=1.68, p_{z0}=0$. The values of all the other parameters are as in Fig. 4b. When $t=100$ time units, the mass of the developed quasar in the core of galaxy 2 is $M_{n2f}=0.04$ and the evolution stops. The value of the Hamiltonian is now $E_J=-2.00051$. In this case, the profile of the LCE clearly indicates that this orbit starts as a chaotic 3D orbit and remains chaotic, during the galactic evolution. In Figure 19c we observe the evolution of the LCE of a 3D orbit with initial conditions: $x_0=2.6764, y_0=0, z_0=0.057, p_{x0}=0, p_{z0}=0$, while the values of all the other parameters are as in Fig. 6b. After time interval of 100 time units the mass of the quasar in the core of host galaxy 2 is $M_{n2f}=0.04$ and the galactic evolution stops. The Hamiltonian settled to the value $E_J=-2.00034$. The profile of the LCE given in Fig. 19c shows that the orbit stars as a regular 3D orbit but after the galactic evolution it becomes a chaotic one. It is evident that, if mass transport was not present, the orbit would have remain regular. The presence of the massive and dense quasar in the core of galaxy 2, has transformed the character of the 3D orbit, from regular to chaotic. Figure 19d depicts the evolution of the LCE of a 3D orbit with initial conditions: $x_0=2.185, y_0=0, z_0=0.044, p_{x0}=0, p_{z0}=0$. The values of all the other parameters are as in Fig. 4b. After time interval of 100 time units the mass of the quasar in the core of galaxy 2 is $M_{n2f}=0.04$ and the galactic evolution stops. The Hamiltonian settled to the value $E_J=-2.00007$. The corresponding profile of the LCE given in Fig. 19d indicates that the orbit stars as a chaotic 3D orbit but after the galactic evolution it becomes a regular one. Therefore, it is evident that, if mass transport was not present, the orbit in this case would have remain chaotic. The presence of the quasar in the core of galaxy 2, has altered the nature of the 3D orbit, from chaotic to regular.

Using the above procedure, we have tested a large number of 3D orbits (approximately 1000) in both time dependent models, describing the formation of massive and dense quasars in the cores of the host galaxies, when mass is transported from their disks. Numerical results, not provided here, suggest that the character of 3D orbits can change either from regular to chaotic and vise versa or not change at all, as the mass is transported and the massive and dense quasars are developed in the central regions of the host galaxies. In particular, from the sample of the 1000 tested orbits in the case of the time-dependent model, we conclude that: 59\% of the orbits altered their nature from regular to chaotic, 23\% remain chaotic, 16\% remain regular and only 2\% changed their character from chaotic to regular. Whether the nature of a 3D orbit will change or not during the galactic evolution described by one of the sets of equations (19), strongly depends on the initial conditions $\left(x_0,p_{x0},z_0\right)$ of each orbit. Moreover, as the change of the value of the Hamiltonian (10) is negligible $\left(\Delta E_J \simeq 10^{-4}\right)$, we can say that the phase space is transformed to itself during the galactic evolution. In order to make this statement more clear would present an example. If we suppose that the evolving time dependent model was describing the 2D system, then as the change of the Hamiltonian is negligible, we could say that in the phase plane of the quiet galaxy 1 shown in Fig. 5b chaotic regions would appear in the central regions and it would be transformed to the phase plane shown in Fig. 5a, as the quasar is formed in the core of host galaxy 1.
\begin{table}
\caption{Radii of resonances when $R=2.5$ and $\Omega_p = 0.417229$, for both active and quiet galaxies 1.}
\centering
\begin{tabular}{|c|c|c||c|c|}
\hline
\textbf{Case} & \textbf{Resonance} & \textbf{Region} & $r_1$ & $r_2$ \bigstrut[t] \\
\hline \hline
\multirow{8}[8]{*}{Active G1} &
\multirow{2}[4]{*}{$\Omega_p = \Omega - 2 \kappa /3$}
          & Direct
                       & 0.0151 & - \bigstrut[t] \\ \cline{3-5}
&         & Retrograde & 0.0147 & 0.2372 \bigstrut[t] \\ \cline{2-5} &
\multirow{2}[4]{*}{$\Omega_p = \Omega - 3 \kappa /4$}
          & Direct
                       & 0.0222 & - \bigstrut[t] \\ \cline{3-5}
&         & Retrograde & 0.0212 & 0.3550\bigstrut[t] \\ \cline{2-5} &
\multirow{2}[4]{*}{$\Omega_p = \Omega - 5 \kappa /8$}
          & Direct
                       & 0.0122 & 0.3882 \bigstrut[t] \\ \cline{3-5}
&         & Retrograde & 0.0119 & - \bigstrut[t] \\ \cline{2-5} &
\multirow{2}[4]{*}{$\Omega_p = \Omega - 7 \kappa /9$}
          & Direct
                       & 0.0255 & - \bigstrut[t] \\ \cline{3-5}
&         & Retrograde & 0.0241 & 0.3905 \bigstrut[t] \\
\hline
\multirow{8}[8]{*}{Quiet G1} &
\multirow{2}[4]{*}{$\Omega_p = \Omega - 2 \kappa /5$}
          & Direct
                       & 0.0273 & - \bigstrut[t] \\ \cline{3-5}
&         & Retrograde & 0.0268 & 0.4731 \bigstrut[t] \\ \cline{2-5} &
\multirow{2}[4]{*}{$\Omega_p = \Omega - 3 \kappa /4$}
          & Direct
                       & 0.0132 & 0.5268 \bigstrut[t] \\ \cline{3-5}
&         & Retrograde & 0.0131 & 0.7156 \bigstrut[t] \\ \cline{2-5} &
\multirow{2}[4]{*}{$\Omega_p = \Omega - 3 \kappa /7$}
          & Direct
                       & 0.0226 & 0.3618 \bigstrut[t] \\ \cline{3-5}
&         & Retrograde & 0.0225 & - \bigstrut[t] \\ \cline{2-5} &
\multirow{2}[4]{*}{$\Omega_p = \Omega - 4 \kappa /9$}
          & Direct
                       & 0.0117 & 0.6054 \bigstrut[t] \\ \cline{3-5}
&         & Retrograde & 0.0115 & - \bigstrut[t] \\
\hline
\end{tabular}
\end{table}

\begin{table}
\caption{Radii of resonances when $R=2.5$ and $\Omega_p = 0.417229$, for both active and quiet galaxies 2.}
\centering
\begin{tabular}{|c|c|c||c|c|}
\hline
\textbf{Case} & \textbf{Resonance} & \textbf{Region} & $r_1$ & $r_2$ \bigstrut[t] \\
\hline \hline
\multirow{8}[8]{*}{Active G2} &
\multirow{2}[4]{*}{$\Omega_p = \Omega - 2 \kappa /3$}
          & Direct
                       & 0.0120 & 0.5529 \bigstrut[t] \\ \cline{3-5}
&         & Retrograde & 0.0117 & 0.1346 \bigstrut[t] \\ \cline{2-5} &
\multirow{2}[4]{*}{$\Omega_p = \Omega - 3 \kappa /4$}
          & Direct
                       & 0.0118 & - \bigstrut[t] \\ \cline{3-5}
&         & Retrograde & 0.0116 & 0.2348 \bigstrut[t] \\ \cline{2-5} &
\multirow{2}[4]{*}{$\Omega_p = \Omega - 5 \kappa /8$}
          & Direct
                       & 0.0097 & 0.6781 \bigstrut[t] \\ \cline{3-5}
&         & Retrograde & 0.0095 & - \bigstrut[t] \\ \cline{2-5} &
\multirow{2}[4]{*}{$\Omega_p = \Omega - 7 \kappa /9$}
          & Direct
                       & 0.0204 & - \bigstrut[t] \\ \cline{3-5}
&         & Retrograde & 0.0193 & 0.2603 \bigstrut[t] \\
\hline
\multirow{8}[8]{*}{Quiet G2} &
\multirow{2}[4]{*}{$\Omega_p = \Omega - 2 \kappa /5$}
          & Direct
                       & 0.0241 & 0.5561 \bigstrut[t] \\ \cline{3-5}
&         & Retrograde & 0.0239 & 0.5926 \bigstrut[t] \\ \cline{2-5} &
\multirow{2}[4]{*}{$\Omega_p = \Omega - 3 \kappa /4$}
          & Direct
                       & 0.0098 & - \bigstrut[t] \\ \cline{3-5}
&         & Retrograde & 0.0097 & 0.7512 \bigstrut[t] \\ \cline{2-5} &
\multirow{2}[4]{*}{$\Omega_p = \Omega - 3 \kappa /7$}
          & Direct
                       & 0.0182 & - \bigstrut[t] \\ \cline{3-5}
&         & Retrograde & 0.0180 & 0.2935 \bigstrut[t] \\ \cline{2-5} &
\multirow{2}[4]{*}{$\Omega_p = \Omega - 4 \kappa /9$}
          & Direct
                       & 0.0087 & 0.4756 \bigstrut[t] \\ \cline{3-5}
&         & Retrograde & 0.0086 & - \bigstrut[t] \\
\hline
\end{tabular}
\end{table}

\section{DISCUSSION}

In the present article, we have constructed a simple 3D gravitational model, in order to study the character of motion in a binary quasar dynamical model. In particular, we present a model for a binary pair of galaxies, following a circular orbit with galactic disks that are aligned with the orbital plane. Each galaxy is assumed to host a massive black hole at its core. The motion of a test particle (star), in the gravitational field of this system is studied and various techniques are used to identify regular and chaotic motion. We have chosen the host galaxies, in our dynamical model, to be disk galaxies due to the fact that the belief that bright quasars reside in massive ellipticals (Dunlop et al., 2003), must be revised as recent observations indicate that the majority of bright quasars are hosted in disk galaxies (Letawe et al., 2006). Observational data regarding radial velocities have shown that the masses of the large spirals are about $2 \times 10^{11}$ solar masses. This mass is consistent with the mass of our disk galaxy 1. As observational data reveal that about $50\%$ of the host galaxies show signs of gravitational interactions, we have decided to present a binary quasar dynamical model of three degrees of freedom, based on a pair of two interacting host galaxies.

A binary system of interacting galaxies hosting quasars in their cores is very complex and therefore, we need to assume some necessary simplifications and assumptions, in order to be able to study the orbital behavior of such a complicated stellar system. Thus, our model is simple and contrived, in order to give us the ability to study different aspects of the dynamical model but nevertheless, contrived models can provide insight into more realistic stellar systems, which unfortunately are very difficult to be studied if we take into account all the astrophysical aspects. Here, we must point out the main restrictions and limitations of our gravitational model. (i) The two host galaxies are assumed to be coplanar and orbiting each other in the same plane on circular orbits. (ii) Our dynamical model only deals with the non-dissipative components of the host galaxies, stars or possibly dark matter particles. It may be that gas accretion would be even more important to the question addressed in the present study. (iii) The potentials we use are rigid and do not respond to the evolving density distribution in a more realistic way. This is because our gravitational model which describes the binary system of the host galaxies is not self-consistent. Self-consistent models are usually deployed when we want to conduct $N$-body simulations. Obviously this is out of the scope of the present paper. Once more, we have to state that the above restrictions and limitations of our model are necessary because otherwise, it would be extremely difficult or even impossible to apply the extensive and detailed dynamical study presented in our study.

In our research, two different cases were investigated: the time independent model and the evolving model, that is the case when mass is transported from the disks of the galaxies to their centers forming a massive and dense quasar in each galactic core. Our numerical calculation indicate that there are several factors responsible for the observed chaotic motion, in the time independent model: (i) the galactic interaction, (ii) the galactic activity, that is the presence of the quasars and (iii) the Lindblad resonances. Furthermore, the presence of the quasars increases the velocities near the central regions of the host galaxies. The value of the velocity depends on the mass of the dense nucleus and the value of its scale length. Regular motion corresponds to low central velocities, while chaotic motion is characterized by high velocities. All the above, strongly indicates that in the centers of active galaxies chaotic motion in high velocities is expected. On the other hand, it was found that the two interacting galaxies, for large values of the distance between their centers $R$, do not present chaotic motion, when the massive quasars are not present in their cores.

As we have mentioned previously, one of the factors responsible for the chaotic motion and other resonance phenomena, such as islandic motion, are several inner Lindblad resonances
\begin{equation}
\Omega_p = \Omega - \frac{n}{m} \kappa \ \ \ ,
\end{equation}
where $\Omega$ and $\kappa$ indicate the circular and the epicycle frequency of the star respectively, while $m$ and $n$ are integers. The main resonances for the two hosts galaxies in both cases (Active and Quiet) together with the corresponding resonance radii $r_1$ and $r_2$, when $R=2.5$ and $\Omega_p = 0.417229$ are given in Tables 2 and 3.

Figure 20a shows a plot of the curves $\Omega-n\kappa/m$ for the active host galaxy 1, as a function of the radius $r$. The numbers 1,2,3 and 4 indicate the curves $\Omega-2\kappa/3, \Omega-3\kappa/4, \Omega-5\kappa/8$ and $\Omega-7\kappa/9$ respectively. The straight lines are the curves $\Omega_p=\pm 0.417229$. The values of all the other parameters are as in Fig. 2. Figure 20b is similar to Fig. 20a but for the active host galaxy 2. As we can see there is a considerable number of resonance radii for both the direct and retrograde orbits in both host galaxies. Details regarding the resonances and the resonance radii can be obtained form Tables 2 and 3. In other words, all the Lindblad resonances given in Tables 2 and 3, are also responsible for the chaotic motion in the two host galaxies. Things are very similar when the quasars are not present in the cores of the two host galaxies. It is also interesting to note that the above resonances produce large chaotic regions for small values of the distance $R$, while for larger values of $R$ (see Fig. 5a-b) the chaotic regions are small, although the resonance radii are still present. This means that in this case the distance between the two galaxies precedence the Lindblad resonances.
\begin{figure}[!tH]
\centering
\resizebox{0.8\hsize}{!}{\rotatebox{0}{\includegraphics*{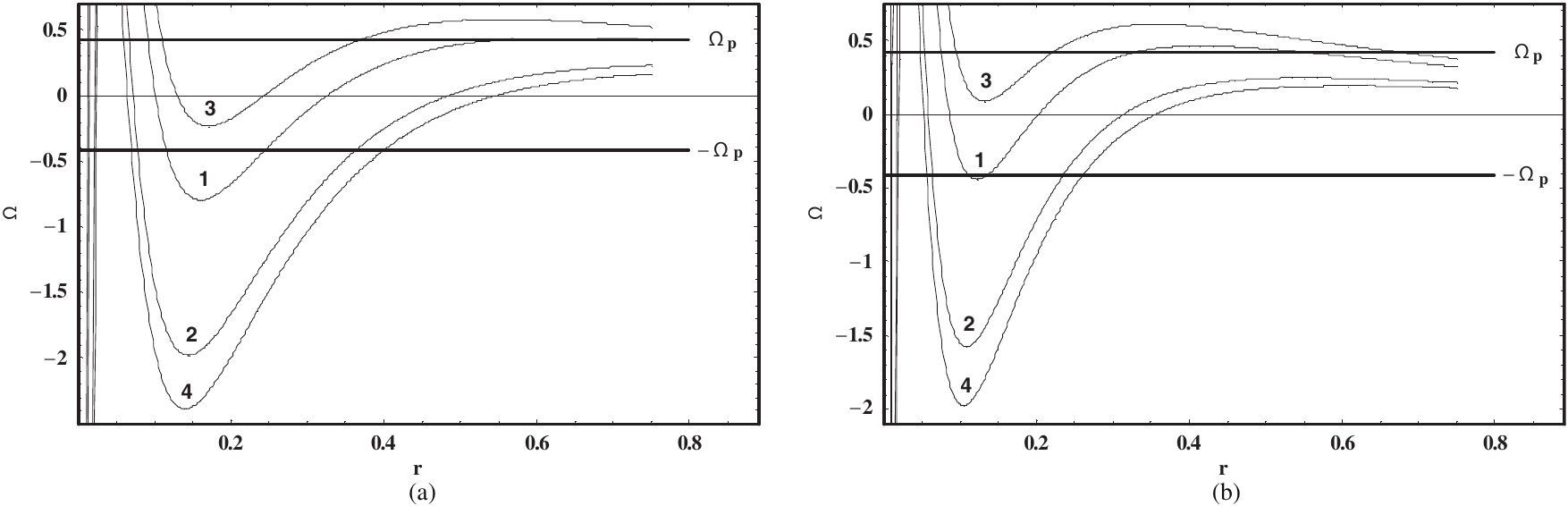}}}
\caption{(a-b): The curves $\Omega-n\kappa/m$ vs $r$, when $R=2.5$ for (a-left): the active host galaxy 1 and (b-right): for the active host galaxy 2. The values of all the other parameters are as in Fig. 2.}
\end{figure}

In order to explore and understand the nature of orbits in the 3D dynamical system, we have used our knowledge obtained from the study of the 2D system. Of particular interest was the determination of the regions of initial conditions in the $\left(x, p_x, z\right), p_y>0, \left(y=p_z=0\right)$ phase space that produce regular or chaotic 3D orbits. As the value of $p_{y0}$ was found from the Jacobi integral (10), we have used the same value of $E_J$, as in the 2D system and took initial conditions $\left(x_0, p_{x0}, z_0\right)$ such as $\left(x_0, p_{x0}\right)$ lies in the chaotic regions of the 2D system. It was found that the motion is chaotic for all permissible values of $z_0$. On the other hand, in the case when $\left(x_0, p_{x0}\right)$ was inside a regular region around the direct and retrograde periodic points, the corresponding 3D orbits are regular for small values of $z_0$, while for larger values of $z_0$ the orbits become chaotic. The particular values of $z_0$ for which the transition from regularity to chaos in 3D orbits is occurred were different for each regular region of the 2D system. Of particular interest are the results given in Table 1, where we define the average minimum value of $z_0$ for each case near the direct and retrograde periodic points, in both host galaxies.

An important role is played by the $L_z$ component of the test particle's angular momentum. It was found that the values of $< L_z >$ for regular 3D orbits are larger, than those corresponding to chaotic 3D orbits. Thus, the $L_z$ component of the angular momentum is a significant dynamical parameter connected with the regular or chaotic character of orbits in both 2D and 3D dynamical systems.

In order to estimate the degree of chaos in the 2D as well as in the 3D dynamical system, we have computed the average value of the maximum Lyapunov Characteristic Exponent - LCE, for a large number of orbits with different initial conditions in the chaotic regions in each case, for a time period of $10^5$ time units. Note that all the calculated LCEs were different on the fifth decimal point in the same chaotic region. The numerical results indicate that the degree of chaos in the 3D binary quasar system is smaller than in similar 2D systems.

It is of great interest to follow the evolution of orbits as the quasars are formed in the central regions of the host disk galaxies. In this procedure, mass is transported from the disks to the nuclei of both galaxies and therefore the quiet galaxies become gradually active, following equations (19). We observe that the final character of the 3D orbits strongly depends on the particular initial conditions $\left(x_0, p_{x0}, z_0\right)$. Therefore, regular orbits can turn to chaotic and vice versa, or maintain their character (ordered or chaotic) during the quasar's formation and since the quasars have been formed. It was observed that a number of regular quasi periodic orbits, starting near the central regions of each host galaxy, become chaotic. This can been seen in a velocity vs. time plot, where the asymmetric profile of the total velocity is in agreement with observational data (see Grosb{\o}l, 2002), where an increase of the stellar velocity is expected, in regions with significant chaoticity. Moreover, observations show that an asymmetric velocity profile indicates chaotic motion.

An interesting question is whether interactions is essential in order to trigger the mass transportation and therefore the galactic activity. This question remains unsolved as observational astronomers have found binary systems with disk galaxies (see Letawe et al., 2006) showing no signs of interaction at all but harboring active nuclei.

Forty years ago, galactic activity and interactions between galaxies were viewed as unusual and rare. Nowadays, they seem to be segments in the life of many galaxies. From the astrophysical point of view, in the present work, we have tried to connect galactic activity and galactic interactions with the nature of orbits (regular or chaotic) and also with the behavior of the velocities of stars in host disk galaxies. We consider the outcomes of the present research, to be an initial effort, in order to explore the dynamical structure of the 3D binary quasar system, in more detail. As results are positive, further investigation will be initiated to study all the available phase space, including orbital eccentricity of the small host galaxy (the galaxy with the smallest value of the total mass) and its inclinations to the primary host galaxy. Moreover, we shall use the outcomes obtained from this initial dynamical study so as to conduct computer $N$-body simulations in a system of interacting galaxies, in order to reveal the change in their orbital properties, throughout merger processes and tidal effects, which are out of the scope of our present work.

\section*{Acknowledgments}

I would like to express my warmest thanks to the anonymous referee for his careful reading of the manuscript and also for his very useful suggestions and comments, which improved greatly the quality and the clarity of the present paper.

\end{document}